%% file: milliscale_techreport.tex
\documentclass[sigconf, nonacm]{acmart}

\usepackage{graphicx}
\usepackage{microtype}
\usepackage{epstopdf}
\usepackage{flushend}
\usepackage{balance}
\usepackage{xcolor}
\usepackage{verbatim}
\usepackage{listings}
\usepackage{booktabs}
\usepackage{nicefrac}
\usepackage{paralist}
\usepackage{subcaption}
\usepackage{algpseudocode}
\usepackage{soul}
\usepackage{subfiles}
\usepackage{multirow}
\usepackage{enumitem}
\usepackage{tabularx}
\usepackage{xspace}

\usepackage{tikz}
\newcommand*\circled[1]{\tikz[baseline=(char.base)]{
            \node[shape=circle,fill=.,inner sep=0pt] (char) {\color{-.}\textsf\footnotesize #1};}}

\newcommand*\squared[1]{\tikz[baseline=(char.base)]{
            \node[shape=rectangle, inner sep=1pt, fill=.] (char) {\color{-.}\textsf\footnotesize #1};}}

\definecolor{comment-red}{rgb}{1,0,0}
\definecolor{OliveGreen}{rgb}{0,0.6,0}
\definecolor{OfficeOrange}{HTML}{ED7D31}

\usepackage{algorithm}
\usepackage{float}
\newfloat{algorithm}{t}{lop}

\newcommand\repourl{https://github.com/sfu-dis/milliscale}

\newcommand{\name}{Milliscale\xspace}

\newcommand{\std}{S3 Standard\xspace}
\newcommand{\express}{S3 Express One Zone\xspace}

\newcommand{\milliscale}{\textsf{Milliscale}\xspace}
\newcommand{\xp}{\textsf{S3X}\xspace}
\newcommand{\sthree}{\textsf{S3 Standard}\xspace}
\newcommand{\ebsg}{\textsf{gp3}\xspace}
\newcommand{\ebsi}{\textsf{io2}\xspace}

\def\thepapertitle{\name: Fast Commit on Low-Latency Object Storage}

\begin{document}
\title{\thepapertitle}

\definecolor{mygreen}{HTML}{03C03C}
\lstset {
language=Python,
mathescape,
basicstyle=\ttfamily\small,
keywordstyle=\ttfamily\bfseries\color{blue},
commentstyle=\color{mygreen},
numbers=left,
numberstyle=\small,
numbersep=5pt,
tabsize=1,
gobble=0,
stepnumber=1,
xleftmargin=15pt,
escapeinside={(@*}{*@)},
morekeywords={},
columns=fullflexible,
}

\settopmatter{authorsperrow=3} 

\setlist[itemize]{leftmargin=*}

\author{Jiatang Zhou}
\affiliation{\institution{Simon Fraser University}}
\email{jiatangz@sfu.ca}

\author{Kaisong Huang}
\affiliation{\institution{University of Calgary}}
\email{kaisong.huang@ucalgary.ca}

\author{Tianzheng Wang}
\affiliation{\institution{Simon Fraser University}}
\email{tzwang@sfu.ca}

\input{0-abstract}
\maketitle

\input{1-introduction}

\input{2-background}

\input{3-overview}

\input{4-design}

\input{5-evaluation}

\input{6-related-work}

\input{7-summary}

\balance
\bibliographystyle{ACM-Reference-Format}
\bibliography{ref}

\end{document}

%% file: 0-abstract.tex
\begin{abstract}
With millisecond-level latency and support for mutable objects, recent low-latency object storage services as represented by Amazon \express have become an attractive option for OLTP engines to directly commit transactions and persist operational data with transparent strong consistency, high durability and high availability. 
But a na\"ive adoption can still lead to high commit latency 
due to idiosyncrasies of \express and modern decentralized logging.  

This paper presents \name, a memory-optimized OLTP engine for low-latency object storage. 
\name optimizes commit latency with new techniques that lowers commit delays and reduces the number of object access requests. 
Our evaluation using representative benchmarks shows that \name delivers much lower commit latency than baselines while sustaining high throughput.  
\end{abstract}

%% file: 1-introduction.tex
\section{Introduction}
\label{sec:intro}
Object storage services (e.g., Amazon S3, Google Cloud Storage and Azure Blob Storage) offer elastic storage, high availability (over 99.9\%~\cite{S3StorageClasses,GCSStorageClasses}) and strong durability (11 nines~\cite{S3StorageClasses}) at low cost~\cite{aws_s3_pricing}. 
They also allow very flexible accesses via HTTP requests. 
These features are highly desirable for database systems, as seen by the wide adoption of object storage for analytics~\cite{Snowflake,DeltaLake,LiquidCache,AnyBlob,DataFusion,DuckDBS3}. 

\begin{figure}[t]
	\centering
	\includegraphics[width=0.65\columnwidth]{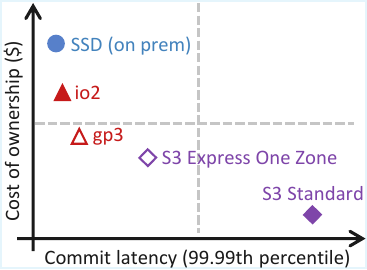}
	\caption{Commit latency vs. cost. 
  Traditional object (\std) and block (local SSD, io2 and gp3) storage incur high commit latency and cost, respectively. 
  } 
	\label{fig:lat-vs-cost}
\end{figure}

Although these features are also highly attractive for OLTP, using object storage as the main storage backend for OLTP engines to \textit{directly persist write-ahead log as objects} is yet to become mainstream. 
A major reason is the high latency of traditional object storage. 
As Figure~\ref{fig:lat-vs-cost} shows, \std exhibits low cost but very high latency. 
As a comparison, block storage services (e.g., Amazon EBS gp3 and io2~\cite{EBS}) allow significantly lower, ms-level commit latency, but come with much higher cost. 
Some also provide lower durability guarantees than S3 (e.g., two 9s in EBS gp3 vs. 11 nines of S3). 
Traditional object storage services also rarely support efficient updates (if any). 
Most existing OLTP services in the cloud thus use proprietary logging services~\cite{Taurus,Socrates}, high-end block storage (e.g., EBS io2) or object storage with SSD caches~\cite{DeltaLake}. 
The downside is they increase system implementation complexity (e.g., by maintaining a separate internal service, or working around the immutability restrictions of \std~\cite{DeltaLake}) and complicate deployment (e.g., by requiring specific compute instances with NVMe SSDs). 

\subsection{Logging on Low-Latency Object Storage}
Recent {low-latency} object storage services as represented by Amazon \express~\cite{S3Express} offer (1) single-digit millisecond-level latency and (2) mutable objects that support append operations. 
These features are a significant improvement over \std's 100ms-level latency and immutable objects, leading to the interesting research question of \textit{would it be feasible for an OLTP engine to transparently have high durability and fast commit by directly logging to an object storage service like \express?}

To answer this question, we set out to adapt ERMIA~\cite{ERMIA}, a memory-optimized OLTP engine, to flush log records to \express. 
While \express enables transparent high availability, high durability and flexible log accesses which can simplify functionality such as hot standby, we observe that transaction commit latency remains high.  
As shown in Figure~\ref{fig:lat-vs-cost}, commit latency under \express can be $\sim$2$\times$ higher with cost approaching gp3's. 
We observe two key reasons below. 

\textbf{Excessive Object Append Requests.}
Modern DBMSs typically use decentralized logging~\cite{NVM-Dlog,TaurusLog,SiloR} to avoid centralized logging bottlenecks~\cite{Aether}. 
The common approach is per-thread logging, which dedicates a private log buffer to each transaction worker thread. 
Log buffer size and the frequency of log buffer flushes directly determine commit latency. 
Using more parallel buffers can help ease the centralized logging bottleneck, 
and flushing more frequently/eagerly with smaller log buffers can help reduce commit latency. 
Both lead to a higher number of log flush (i.e., object append) requests, which in turn lead to more network roundtrips needed for flushing log records. 
Some modern memory-optimized DBMSs also tend to store full records in the log~\cite{QueryFresh,Silo} (rather than the potentially much smaller delta) to simplify implementation. 
This further increases log data volume, again increasing data transfer latency on the commit path. 
Moreover, cloud object storage services use pay-as-you-go~\cite{aws_s3_pricing} pricing models that charge by data volume, storage duration \textit{and} the number of object access requests. 
As a result, higher number of requests also leads to higher cost. 

\textbf{Commit Dependencies.} 
Under decentralized logging,   
to commit a transaction $T$, the OLTP engine must ensure $T$'s own \textit{and} depending log records which may be generated by other threads in their private log buffers~\cite{CowBook} have been persisted.
Compared to using a single log buffer, this naturally adds commit latency by potentially requiring (1) flushing multiple log buffers and (2) dependency tracking~\cite{AetherVLDBJ,AutonomousCommit}. 
That is, a transaction $T$ may read or overwrite the result of another transaction $S$ recorded in another log. 
Committing $T$ requires both threads' log buffers be flushed, adding commit latency. 
Worse, $T$ may have to wait for the persistence of log buffers which it does not depend on, because its serialization order is later. 
Such false positive dependencies can lead to further higher commit latency.

\subsection{\name}
This paper presents \name, a memory-optimized OLTP engine that offers fast commit over low-latency, mutable object storage in the cloud. 
The core of \name is a new decentralized logging protocol 
with optimizations that respectively reduce excessive log flush requests and unnecessary dependency delays. 

We make two observations as we explore ways to reduce object append requests. 
(1) The frequency of log flush is directly affected by the speed the log buffer is filled, which in turn is determined by the amount of concurrent requests targeting each buffer. 
(2) On modern CPUs contention on a single log buffer by a small number of threads (e.g., 2--4) within a socket is negligible. 
Based on these observations, we depart from the common per-thread logging and propose \textit{restricted decentralized logging}: 
Instead of assigning each thread a log buffer, we assign each group of (e.g., two) threads a proportionally larger (e.g., 2$\times$) log buffer. 
As Section~\ref{sec:design} elaborates, by carefully selecting log buffer size based on the performance characteristics of object storage, restricted decentralized logging reduces object append requests, yet without affecting commit latency. 

On top of restricted decentralized logging, \name reduces delays caused by false positive dependencies with a \textit{lightweight record-level dependency tracking} approach. 
Instead of tracking at the page level~\cite{RethinkLog,NVM-Dlog} or global durable commit timestamps~\cite{LeanStoreEvolve},  
Each transaction $T$ tracks the commit timestamp of its most recent direct predecessor $P$ that generated the record read or updated by $T$. 
Once such $P$ has committed, $T$ can safely commit. 
This is done by keeping a transaction-local variable that is updated while reading/update records. 
The process is lightweight and only requires obtaining the commit timestamp on each version, which is usually already maintained by the underlying concurrency control protocols. 
With fine-grained record-level tracking, \name avoids a large number of false positive dependencies that cause unnecessary delays. 

Note that our goal is \textit{not} to achieve the lowest possible commit latency and compete with NVMe SSD based solutions, but to show the potential for modern object storage of delivering satisfactory commit latency and throughput. 
We thus do not involve intermediary layers such as NVMe SSDs that are ephemeral and only available on certain compute instance types. 
Overall, the commit latency obtained by \name is still higher than using NVMe SSDs. 
However, we believe this is a meaningful step towards cloud-native OLTP services with simpler system architectures but rich durability and availability features.   

We built \name on top of the aforementioned OLTP engine, ERMIA~\cite{ERMIA}, whose logging subsystem was designed for block storage.  
\name inherits ERMIA's memory-centric optimizations (e.g., indexing and concurrency control), but replaces ERMIA's logging subsystem with our new decentralized logging approach over low-latency, mutable object storage. 
Our evaluation using YCSB~\cite{YCSB} microbenchmarks and TPC-C~\cite{TPCC} on a 32-vCPU Amazon EC2 \textsf{c6in.8xlarge} instance and \express shows that \name delivers competitive throughput and much lower (up to 51.9\%) tail latency when compared to baselines. 
In particular, \name can bring the 99.99 percentile commit latency for write-intensive workloads of the unoptimized \express to the level that is similar to gp3's, with the same elasticity and consistency guarantees as \std's. 

Techniques in \name are applicable to other object storage services. 
Restricted decentralized logging can be useful beyond logging that involve writing out buffers of data to storage (e.g., to reduce the number I/O requests in a busy system). 
One may further optimize the read path by adding SSD-based caches~\cite{LiquidCache,DeltaLake}. 
We leave it as future work to explore these directions. 

\subsection{Contributions and Paper Organization}
We make four contributions.
\circled{1} We revisit the idea of OLTP over object storage and make the case for directly persisting write-ahead log on low-latency object storage. 
\circled{2} We propose a new decentralized logging approach that jointly optimizes throughput and commit latency. 
\circled{3} Based on the proposed techniques, we build \name, the first memory-optimized OLTP engine that directly commits to low-latency object storage. 
\circled{4} We provide a comprehensive evaluation of \name and baselines to validate our design decisions. 
\name is open-source at \url{\repourl}.

Next, Section~\ref{sec:bg} gives the necessary background on cloud storage and characterizes \express to motivate our work. 
Sections~\ref{sec:principles}--\ref{sec:eval} cover the design and evaluation of \name. 
We discuss related work in Section~\ref{sec:related} and summarize in Section~\ref{sec:summary}.

%% file: 2-background.tex
\section{Background}
\label{sec:bg}
In this section, we give the necessary background on low-latency object storage and logging to motivate our work. 

\subsection{Object Storage Basics}
Object storage treats data items as ``objects'' which are identified by unique keys and accessed via HTTP requests, instead of file-based POSIX~\cite{POSIX} APIs. 
An object is a byte container and multiple objects can be grouped together to be stored in a bucket.  
For example, Amazon S3 exposes \texttt{GetObject}/\texttt{PutObject} APIs that take a key as an input (along with other parameters, if any) and perform object read/write operations.
Other APIs for operations such as delete and metadata queries are also provided. 
Requests can be issued from any compute client via data center networks or the internet, but most high-performance deployments use the former to lower latency, which we assume in this work. 

Most cloud vendors offer object storage services with strong consistency and abundant bandwidth thanks to modern data center networking~\cite{AnyBlob}. 
Although these services differ in their internal designs, they typically adopt S3 APIs which have become the de facto standard. 
We thus assume Amazon S3 in this paper and leave it as future work to explore other services. 

Traditional object storage services (e.g., \std) are characterized as offering strong durability (e.g., 11 nines), strong consistency, high availability (over 99.9\%) and low cost. 
The downside, however, is that they exhibit high access latency (100ms level) and objects are immutable (i.e., no update allowed once written). 
These have limited the use of traditional object storage mainly to read-dominant workloads such as analytics~\cite{AnyBlob,LiquidCache}.   
Although some systems~\cite{DeltaLake,Lakebase} extensively use object storage for operational data, they need complex techniques to workaround \std's immutability and rely on aggressive caching using local NVMe SSDs that are only available on high-end compute instances. 

\subsection{\express: S3 After 20 Years}
\label{subsec:s3xp}
Low-latency object storage services such as Amazon \express improve traditional services with single-digit, millisecond-level latency and mutable objects via append operations. 

\textbf{Object Operations.}
\express also exposes the HTTP-based APIs such as \texttt{GetObject}/\texttt{PutObject}. 
Append operations are issued using \texttt{PutObject} by specifying a write offset (\texttt{writeOffsetByte} in AWS S3 SDK) that is the end of the object. 
AWS's current implementation allows an object to be appended for up to 10,000 times~\cite{S3SDK}. 
Each append operation creates a ``part'' in the object and the maximum size per append (i.e., one part) is 5GB. 

\textbf{Performance.}
We issue S3 requests from an AWS EC2 instance to evaluate the latency of \texttt{GetObject} and \texttt{PutObject} (detailed setup in Section~\ref{sec:eval}). 
Full-object writes behave similarly to append operations, so we omit their results and focus on append operations which are the most relevant to our use case (write-ahead logging). 

\begin{figure}[t]
\centering
\includegraphics[width=\columnwidth]{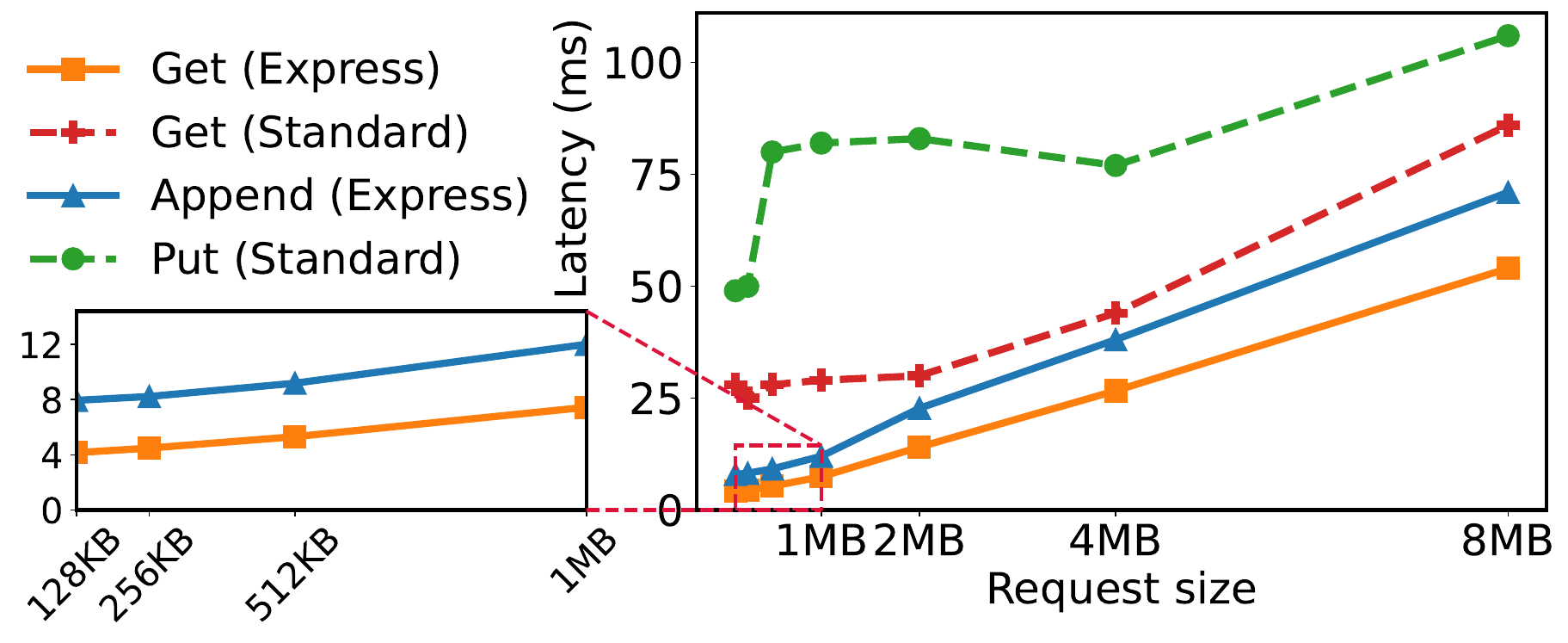}
\caption{Object get, append, and put latency (ms) of \std vs. \express under varying request sizes.
\express can offer single-digit ms latency at small request sizes (e.g., 512KB), making it a good fit for OLTP.}
\label{fig:microbench}
\end{figure}

As Figure~\ref{fig:microbench} shows, \express gives similar append latency ($\sim$8ms) for request sizes of 128--512KB.  
Beyond 512KB, append latency grows linearly. 
At 2MB request size, \std exhibits much higher latency of $\sim$77ms, while the number for \express is $\sim$71\% lower ($\sim$22ms). 
This means 512KB is the minimum request size to avoid excessive networking penalty. 
Since an object can accept at most 10,000 appends, the maximum object size is 4.88GB with 512KB request size. 
The DBMS must be able to segment the log into individual objects, which we describe later. 

Retrieving data using \texttt{GetObject} is generally faster than writing for both \std and \express. 
The latter exhibits $\sim$5ms of latency for all request sizes of up to 256KB as shown in Figure~\ref{fig:microbench}. 
Reducing request size below this threshold yields no further benefit; 
increasing it to 1MB only marginally raises latency. 
Beyond 2MB request sizes, read latency starts to grow linearly. 

These results indicate that the ideal request size for \express is between 256KB to 1MB, potentially leading to an upper-bound latency of $\sim$10ms for committing OLTP transactions, which is only 1/10th of what would be incurred using \std. 
We discuss how \name leverages this observation later.

\textbf{Availability and Durability.}
Compared to \std which replicates data across three AZs, 
\express stores data in a single AZ to lower latency. 
The downside is lower availability compared to \std, leaving it to the DBMS that is built on top of \express to mitigate. 
A straightforward solution is for the OLTP engine to replicate objects to additional S3 buckets in other AZs or regions. 

\textbf{Pricing.}
\express employs a pay-as-you-go pricing model to charge based on actual usage. 
This includes (1) storage cost which is charged based on the amount of and for how long the data is stored, 
(2) number of object access requests (e.g., pay per \texttt{PutObject}), 
and (3) bandwidth usage between the client (e.g., an EC2 instance) and S3. 
While storage cost is only related to data size and duration (similar to how other cloud storage), the other two items are closely related to the number of requests. 
As of early 2026, AWS charges \$1.13 per million requests. 
On top of that, each GB of data uploaded to \express costs \$0.0032. 
The amount of data uploaded is in turn determined jointly by \textit{the number of upload requests $\times$ request size}. 
For example, uploading 2GB of data by issuing 1 million append operations (i.e., 2MB or 0.002GB per request) will cost 
\textsf{\$1.13 + 1,000,000 requests $\times$ 0.002GB $\times$ \$0.0032.}
The \$1.13 is for issuing the 1 million requests alone, and the remaining is for uploading the 2GB of data.  
As a result, reducing the number of S3 requests (which is a major goal of our work) would also help lowering cost. 

\subsection{Why OLTP over \express?}
At first glance, \express's pricing model is not friendly to OLTP which is characterized as having high write frequency. 
However, given the desirable features of S3 and drawbacks of other alternatives, we argue \express can be a desirable choice for a few reasons. 

First, services like \express now provide the aforementioned strong consistency, high availability and strong durability, all \textit{transparently}. 
This can greatly simplify the making of cloud-native database systems which have been mostly relying customized logging services maintained by teams internal to the DBMS vendor, such as the PLOG in Huawei Taurus~\cite{Taurus} and XLOG in Microsoft Socrates~\cite{Socrates}. 
Often times, without such proprietary services, block storage services like EBS are used, yet they do not always provide the same level of guarantees. 

Second, \express inherits the same flexible data access via HTTP requests from its standard counterpart. 
Objects are directly accessible by multiple compute nodes, simplifying hot standby setups where backup servers can directly read from the objects generated by the primary node.  
As a comparison, typical block services like EBS gp3 do not support accessing a volume at the same time by multiple instances, or have limitations (e.g., io2's ``multi-attach'' feature allows multiple servers to mount the same volume, but does not allow them to access the data simultaneously under popular file systems like ext4 and XFS~\cite{io2MultiAttach}). 
Without \express, DBMS builders have to spend extra effort and employ additional hardware to deliver the aforementioned features. 
Such extra human capital cost is not typically taken into consideration when comparing service cost.  

Finally, log space is usually bounded and reused with periodic checkpoints. 
This means the storage cost mentioned in Section~\ref{subsec:s3xp} will stabilize instead of growing as data size grows, leaving the number of requests and sheer amount of data being transferred per roundtrip the only metrics to optimize for. 
Optimizing for these metrics is closely related to how modern write-ahead logging works, which we describe next. 

\subsection{Modern Decentralized Logging}
\label{subsec:log}
Traditional write-ahead logging is centralized:
all worker threads compete to insert into a single log buffer, which then gets flushed periodically to group commit transactions. 
Modern OLTP engines have largely moved away from this model to avoid contention on the centralized log buffer~\cite{AetherVLDBJ}. 
A common approach is decentralized, per-thread logging where each worker thread is allocated a private log buffer~\cite{SiloR,Silo,TaurusLog,NVM-Dlog}, eliminating log buffer contention. 
Each log buffer is periodically flushed (e.g., upon group commit or time out) to stable storage, by appending to a file or object in S3. 

\textbf{Pipelined Group Commit.}
On top of decentralized logs, some systems use pipelined group commit~\cite{Aether} to improve throughput; 
we assume and design \name with such optimization. 
With pipelined group commit, a transaction is detached from the worker thread once it ``pre-committed,'' i.e., has finished processing all logic and inserted all log records to a log buffer. 
The transaction is then placed on a commit queue to wait for final commit. 
Meanwhile, the thread that was serving $T$ can continue to serve another transaction. 
A background thread monitors the currently durable log sequence numbers (LSNs) and dequeues transactions and notify clients once their log records have been persisted. 

\textbf{Commit Dependencies.}
Pipelined group commit frees worker threads from being blocked by log flush I/O, but can form dependencies between transactions that require additional care upon commit. 
Consider a transaction $T$ that updated record $x$ and has just pre-committed, making $T$'s results immediately visible. 
Any transaction $S$ that accesses $x$ will form a dependency on $T$: 
committing $S$ requires log records generated by $T$ be persisted. 
This was trivial in centralized logging as all log records are inserted to the only system-wide log buffer, flushing which will implicitly persist all depending log records~\cite{Aether}. 
Under decentralized logging, however, this often mandates flushing multiple log buffers.  
In the example, to commit $S$, the engine would require both log buffers serving two threads be flushed. 
Tracking exact dependencies (e.g., by maintaining a centralized dependency graph) can further incur performance bottlenecks~\cite{AetherVLDBJ}. 
Therefore, various prior approximations have been proposed.
They typically involve maintaining global log sequence numbers (LSNs) and allow only transactions with commit LSN lower than the lowest durable LSN among all logs to commit~\cite{NVM-Dlog,RethinkLog,BorderCollie}. 
We elaborate the issues of these approaches while introducing \name's solution later.

%% file: 3-overview.tex
\section{Design Principles}
\label{sec:principles}
The characteristics of \express lead to the following design guidelines for \name: 
\begin{itemize}
\item \textbf{Frugal on Object Operations.}
The size and number of requests of object operations (especially \texttt{PubObject} for logging) must be carefully calibrated and planned to reduce 
data transfer delays. 

\item \textbf{Short Dependency-Induced Delay.}
The raw latency of \express can be as low as single-digit millisecond. 
\name should try to keep the end-to-end commit latency close to it by cutting unnecessary software-induced delays. 

\item \textbf{Low Latency while Sustaining High Throughput.}
While our focus is to reduce (tail) commit latency, it is desirable to maintain high throughput as achieved by prior work. 
\end{itemize} 

\section{\name Design}
\label{sec:design}
In this section, we first give an overview of \name and then describe its design in detail. 

\subsection{Overview}

\begin{figure}[t]
	\centering
	\includegraphics[width=\columnwidth]{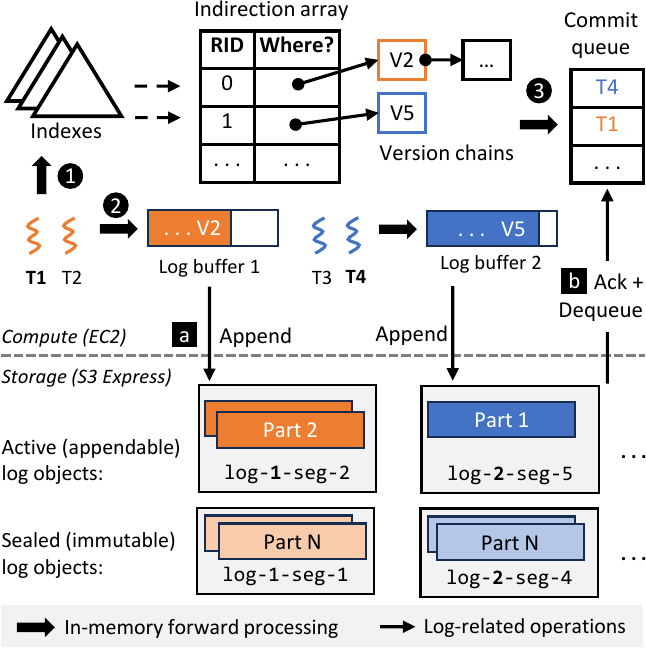}
	\caption{\name Overview.
  Worker threads access records via indexes and version chains. 
  The log is segmented into objects which are sealed once reaching append limitation. 
  To reduce S3 requests while retaining high throughput and low commit latency, threads are grouped to share log buffers. 
  Pre-committed transactions stay in a commit queue until their own and depending log records are persisted. 
}
	\label{fig:arch}
\end{figure}

\name is an OLTP engine with a focus on reducing transaction commit latency over low-latency, mutable object storage. 
Like most cloud-native DBMSs, \name disaggregates compute and storage. 
Transaction requests are accepted on a compute node (e.g., an Amazon EC2 instance) which employs worker threads to access data and write out log records to object storage. 
We aim to optimize the commit path, so we focus on the logging subsystem and adopt representative memory-oriented optimizations from ERMIA~\cite{ERMIA} for multi-versioned concurrency control, indexing and data organization. 
This way, \name maintains ACID properties while leveraging in-memory optimizations and object storage. 

\textbf{Compute Layer (Forward Processing).}
Records in \name are identified by logical record IDs (RIDs). 
Versions of the same records are linked in a version chain in new-to-old order~\cite{MVCCEval}.  
Each table is represented by an indirection array that maps each RID to the latest version of the record.  
Each version is stamped with by commit timestamp which is the commit sequence number (CSN) of the transaction that created it.  
CSNs are drawn from a central counter using the atomic fetch-and-add instruction~\cite{IntelManual} when the transaction entering pre-commit. 
In addition, each transaction is associated with a begin timestamp obtained by reading the global CSN counter upon creation or reading the first record. 
As shown in Figure~\ref{fig:arch}(top), \circled{1} transactions traverse indexes and version chains to access data, and \circled{2} insert into log buffers as they modify the database. 
On top of decentralized logging, \name allows limited sharing of log buffers among worker threads (but without introducing a bottleneck) and writing only the modified fields of each record. 
These designs collectively reduce S3 append requests. 

\textbf{Storage Layer (Commit Path).}
Following classic pipelined commit~\cite{Aether}, \circled{3} after a transaction $T$ finishes forward processing, the worker thread pre-commits $T$ by placing it onto the commit queue. 
The thread then continues to handle the next request, while $T$ waits for its own and depending log records to be persisted in S3. 

\name uses \express as the main persistent home of data to benefit from S3's high durability, scalability and low-latency append operations. 
Importantly, \name does not rely on intermediate levels such as NVMe SSDs on high-end compute nodes, allowing it to be deployed on low-cost compute instances. 
\squared{a} Log buffers are written to \express periodically using S3 append operations when they are full or reach a timeout. 
As shown in Figure~\ref{fig:arch}(bottom), the log is segmented into objects, each of which consists of parts appended to it as a result of log buffer flushes. 
In \express, \name maintains an active object per log and each log buffer flush appends a part. 
For example, in Figure~\ref{fig:arch}(bottom), \textsf{log-1} is shared by $T1$ and $T2$, and currently has two parts in its active object.  
Meanwhile, $T1$ has inserted its new version $V2$ into \textsf{Log buffer 1} and pre-committed. 
After \textsf{Log buffer 1} has been filled or a timeout happens, it will be appended to the currently active object (\textsf{log-1-seg-1}) as the third part.  
The active object is sealed (immutable) once it reaches the append limit (10,000 times as mentioned in Section~\ref{subsec:s3xp}). 
For example, \textsf{log-1} in Figure~\ref{fig:arch}(bottom) has one sealed log segment (\textsf{log-1-seg-1}), whereas \textsf{log-2} (shared by $T3$ and $T4$) has four sealed segments (\textsf{log-2-seg-1} -- \textsf{log-2-seg-4}).
\squared{b} Upon each successful append, we examine the commit queue to release transactions whose own and depending log records are all persisted. 

Overall, commit latency is determined by for how long a transaction remains pre-committed, which is determined by (1) the size and frequency of log buffer flushes, and (2) dependency tracking delays. 
In the rest of this section, we describe \name's approaches to reducing both kinds of delays, while maintaining high throughput.

%% file: 4-design.tex
\subsection{Restricted Decentralized Logging}
\label{subsec:rdl}
Traditional decentralized logging dedicates each worker thread a log buffer to completely eliminate log buffer contention. 
However, it mandates more log buffer flushes. 
The same amount of work done by a group of parallel threads (generating the same amount of log data as using one log buffer) is now distributed to multiple log buffers, each of which needs to be persisted upon commit. 
Recent NVMe SSDs have further enabled such designs with high bandwidth and low latency~\cite{NVMeStorageEngines}; 
it is often desirable to parallelize as much as possible with per-thread logs. 
On object storage, however, this is not always the case: 
using more log buffers increases both the number of S3 append requests and stragglers due to networking conditions, leading to lower performance. 

\textbf{Per-Group Logging.}
We observe the key culprit is the high number of log buffers in modern multicore systems, reducing which would naturally mitigate the problem. 
Therefore, instead of adopting traditional per-thread logging, \name takes a middle ground to propose \textit{restricted decentralized logging} which makes the degree of log decentralization adjustable. 
Restricted decentralized logging employs per-thread-group logging where worker threads are divided into groups, each of which is allocated a separate log buffer. 
Threads within a group will share and contend for the same log buffer (e.g., two threads per log in Figure~\ref{fig:arch}). 
As Section~\ref{sec:eval} shows, however, such contention has negligible performance impact when concurrency (i.e., group size) is carefully controlled. 
That is, the log is decentralized across a set of thread groups, whose size is adjustable depending on the tolerable contention level on the given hardware. 
This allows us to reduce the total number of log buffers in the system, reducing the number of S3 appends and impact of stragglers on performance (i.e., improving tail latency). 

\begin{figure}[t]
	\centering
	\includegraphics[width=\columnwidth]{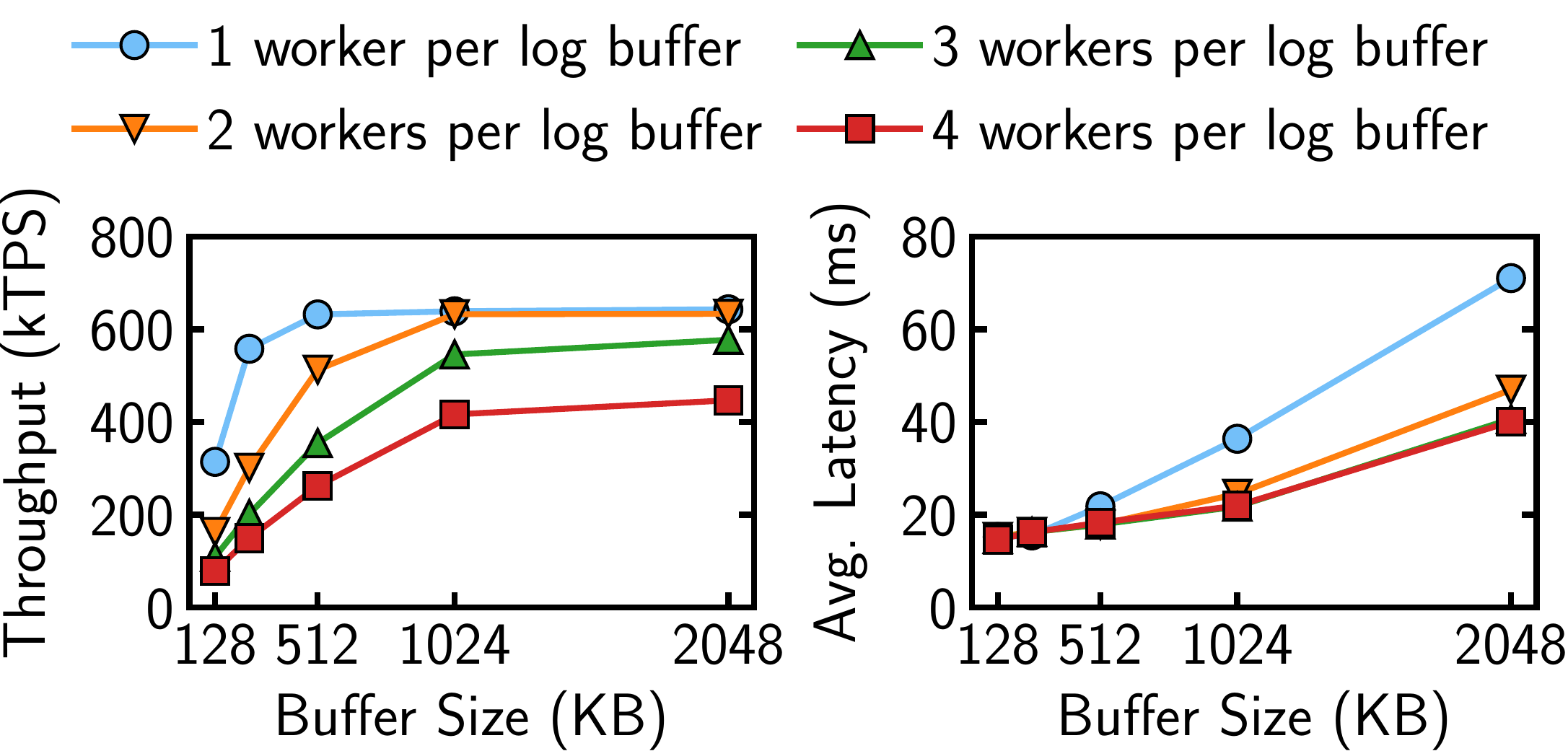}
	\caption{Throughput (left) and average commit latency (right) of TPC-C under different log buffer sizes and sharing ratios. 
  S3 latency dominate with a 4:1 ratio, whereas a 2:1 ratio leads to the highest throughput and lowest latency.}
	\label{fig:tpcc_combine1v2v4}
\end{figure}

\textbf{Proportional Log Buffer and Thread Group Scaling.} 
While the idea of sharing a log buffer per group of thread is straightforward, for it to be truly effective, one should carefully choose (1) group size (i.e., thread-to-log sharing ratio) and (2) log buffer size, as  
they collectively determine both commit latency and the number of requests to object storage. 
Compared to per-thread logging, log buffer size should be adjusted so that the average amount of log buffer space per worker thread is at least the log buffer size as in per-thread logging. 
This is necessary to guarantee meaningful reduction of flushes but also means using larger log buffers, while maintaining commit latency comparable to that of using per-thread logging. 
However, with larger log buffers (hence higher sharing ratio), their fill rate drops (i.e., it takes longer to fill out each log buffer until full). 
This leads to longer transaction commit latency under pipelined group commit which is necessary for achieving high throughput (Section~\ref{subsec:log}). 
For example, suppose a transaction $T$ has pre-committed as the first to insert to the log buffer in its thread group. 
$T$ is then placed on the commit queue until the log buffer is full or a time out happens. 
In a busy system with constant activity, the former would be the usual case and $T$ would be waiting on the commit queue until the log buffer is full and persisted with one S3 append request. 
The larger the log buffer is, the longer $T$ needs to wait for other transactions to finish filling out the log buffer before $T$ can be fully committed (i.e., removed from the commit queue). 

Therefore, it is desirable to \textit{proportionally} increase the log buffer size by group size. 
For example, with a group size of two (i.e., two threads sharing one log buffer), the log buffer size should be 2$\times$. 

\textbf{Log Buffer Size vs. Transfer Latency.}
Increasing log buffer size proportional to group size mitigates the impact of lower log buffer fill rates, but can worsen log buffer flush delay since each request now is larger. 
\name leverages the performance characteristics of \express to solve this problem. 
As we have elaborated in Section~\ref{subsec:s3xp}, the latency to finish an append request only starts to grow linearly with request size beyond 1MB. 
Thus, it takes a similar amount of time to append a 512KB and a 1 MB log buffer. 
Consequently, setting the log buffer size to 1MB with a sharing ratio of two would lead to the commit latency equivalent to using per-thread logging with 512KB log buffers. 

Figure~\ref{fig:tpcc_combine1v2v4} illustrates the tradeoff space of sharing ratio and log buffer size for throughput (left) and latency (right) under TPC-C (detailed setup in Section~\ref{sec:eval}). 
Under per-thread logging (i.e., 1 worker per log buffer in the figure), throughput plateaus at 512KB log buffer. 
With larger log buffers (e.g., 1MB/2MB), a sharing ratio of two keeps the same throughput as per-thread logging because 
each thread on average has 512KB/1MB of log buffer space. 
Yet a sharing ratio that is too high (e.g., 4) would reduce the average log buffer space per thread, increase contention and flush frequency, eventually leading to lower throughput. 
Figure~\ref{fig:tpcc_combine1v2v4}(right) shows the corresponding average commit latency. 
Different from Figure~\ref{fig:tpcc_combine1v2v4}(left), a higher sharing ratio accelerates log buffer fill rate, leading to lower commit latency. 
Meanwhile, it is notable that under 1MB buffer size, a sharing ratio of two or four exhibits similar latency profile to using 512KB log buffers without any sharing.  
With dual-goal of maintaining high throughput (same as using per-thread logging or 1 worker per log buffer in the figure) and low commit latency, we therefore choose a sharing ratio of two and 1MB log buffer size. 
As our evaluation in Section~\ref{sec:eval} shows, such choice of sharing ratio and log buffer size is applicable to wide range of OLTP workload patterns, and can significantly reduce tail latency. 

\subsection{Record-Level Dependency Tracking}
\label{subsec:dep}
\begin{figure}[t]
	\centering
	\includegraphics[width=0.85\columnwidth]{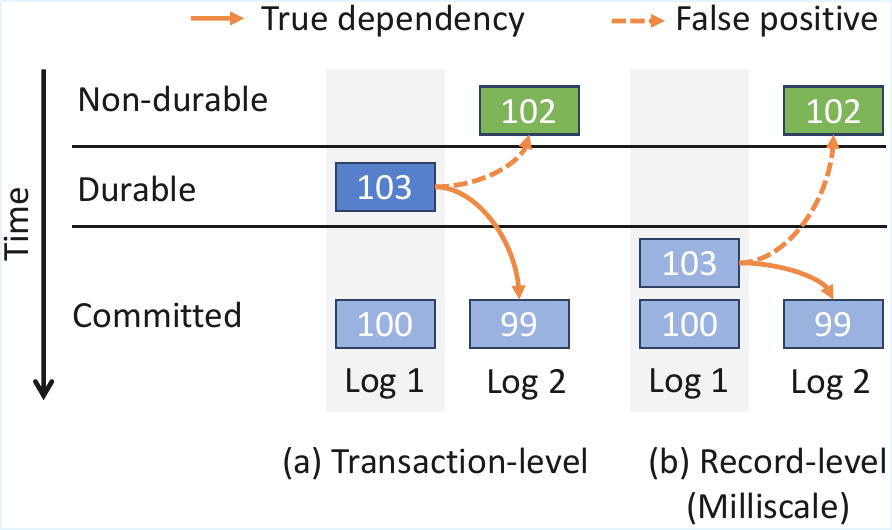}
	\caption{Transaction (a) vs. record (b) level dependency tracking which. 
  Tracking at the record-level allows transaction with CSN 103 to commit, by skipping waiting for the transaction with CSN 102 to commit first.} 
	\label{fig:false-dep}
\end{figure}

Restricted decentralized logging reduces log flushes and log data volume, but still leaves additional software-induced delays compared to centralized logging (Section~\ref{subsec:log}). 
To commit a transaction $T$, one needs to ensure both $T$'s own log records and any log records $T$ depends on are both persisted. 
This requires tracking dependencies between transactions and only fully commit transactions (i.e., notify the application) that meet both requirements.   
\textbf{False Positive Dependencies.}
Recording detailed, accurate dependency information (e.g., using dependency graphs) requires both additional log space and CPU cycles, 
so most systems approximate dependencies among transactions. 
Earlier work~\cite{NVM-Dlog,RethinkLog} uses LSNs and GSNs to establish the partial order between transactions. 
However, these approaches are page-based that will incur false positive dependencies for transactions that access different records (thus no dependency) on the same page. 
More recent approaches employ more fine-grained, transaction-level dependency tracking~\cite{LeanStoreEvolve,BorderCollie} avoid such false positive dependencies. 
Specifically, a transaction can only commit when its begin timestamp is lower than the lowest durable timestamp across all logs. 
However, we observe that this still can lead to false positives that significantly affect commit latency. 
Figure~\ref{fig:false-dep}(a) shows an example with two logs, each dedicated to a group of threads (or a single thread under per-thread logging). 
The transactions are represented in boxes with their begin timestamp (or commit timestamps, depending on the system's concurrency control protocol); 
for example, we refer to the transaction with timestamp 103 as $T103$. 
Here, the lowest non-durable timestamp across all logs is 102, allowing any transactions below it (i.e., $T99$ and $T100$) to commit. 
Meanwhile, $T103$ has pre-committed with its log records fully persisted. 
However, since it is serialized after T102 (e.g., by acquiring its timestamp slightly later using atomic fetch-add), even though $T103$ does not depend on $T102$, it still cannot commit before $T102$ does, leading a false positive dependency. 

\begin{figure}[t]
	\centering
	\includegraphics[width=0.8\columnwidth]{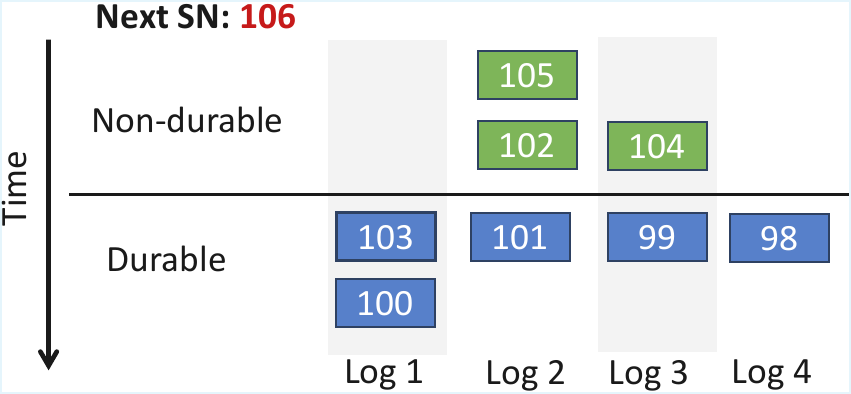}
	\caption{\name takes the minimum non-durable CSN as the gCSN. 
    If a log (Log 1 and Log 4) does not have a non-durable transaction, then its non-durable sequence number is considered as the next possible sequence number (106), i.e., gCSN = min(106, 102, 104, 106) = 102.}
	\label{fig:compute-gCSN}
\end{figure}

\textbf{Record-Level Dependency Tracking.}
\name takes a step further to alleviate this problem with more fine-grained record-level dependency tracking. 
During forward processing, each transaction maintains the commit timestamp of its most recent direct predecessor that it depends on (called maximum dependency sequence number, or DSN). 
For example, in Figure~\ref{fig:false-dep}(b), if $T103$ reads or update a version produced by $T99$, it will update its DSN to be 99 if its current DSN is lower than 99. 
As log buffers get flushed, the engine computes a system-wide commit sequence number based on each log's durable and non-durable sequence numbers. 
Specifically, \name will check all non-durable sequence numbers in each log and take the minimum as the commit sequence number, as shown in Figure~\ref{fig:compute-gCSN}. 
The pipelined group commit thread then can safely commit any transaction whose DSN is lower than the commit sequence number as this indicates that all the transaction's dependent transactions have committed.
Compared to traditional transaction-level tracking, as shown in Figure~\ref{fig:false-dep}(b), record-level tracking allows $T103$ to commit by avoiding a false positive dependency on $T103$.

\begin{figure}[t]
  \centering
  \includegraphics[width=0.8\columnwidth]{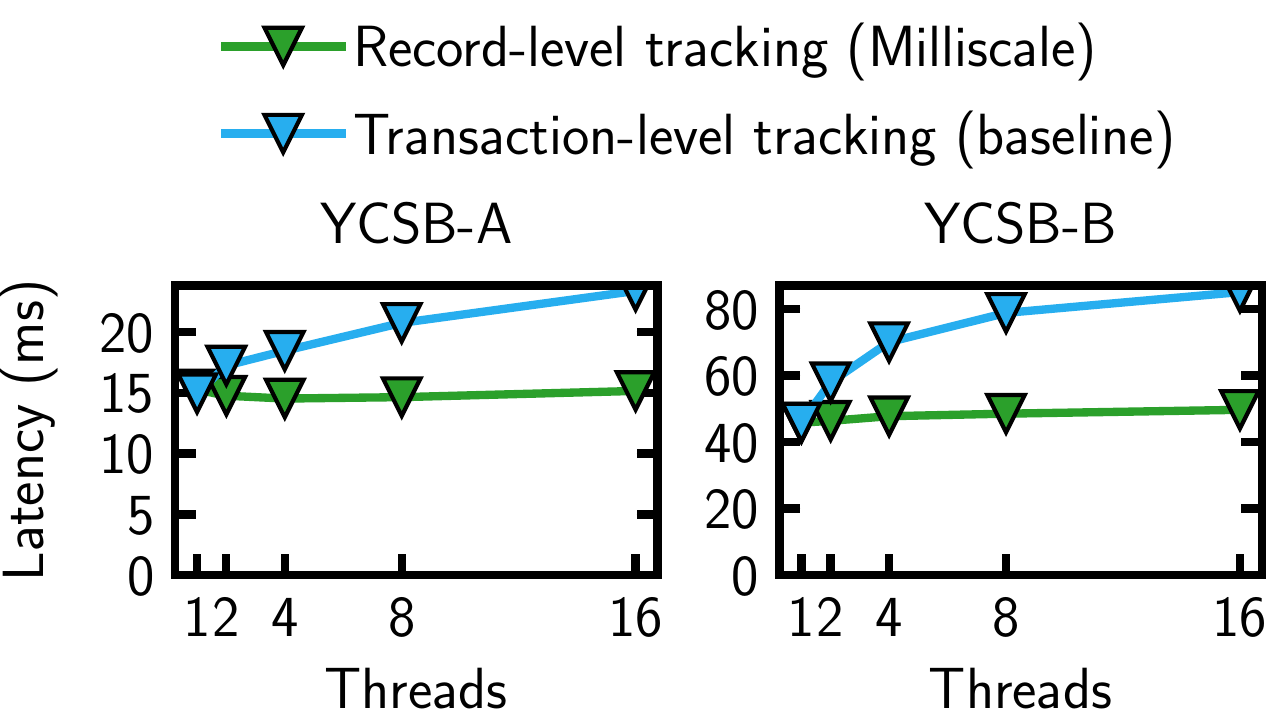}
  \caption{Average commit latency of YCSB-A (left) and YCSB-B (right) under uniform distribution. 
  \name avoids more false positive dependencies by tracking at the finer granularity of records than transaction-level tracking (baseline).}
  \label{fig:avg_latency}
\end{figure}

Maintaining a DSN per transaction is lightweight. 
Since it is transaction-local, no additional concurrency control is required. 
Under MVCC (which \name is based on), each version already carries the commit timestamp of its creator. 
The impact on recovery is also small, which we will discuss later. 
As shown in Figure~\ref{fig:avg_latency}, the average commit latency of YCSB-A (50\%/50\% read/update) and YCSB-B (95\%/5\% read/update) transactions under uniform distribution drops by up to over 35\% when compared to transaction-level tracking. 
We further explore the effect of record-level dependency tracking on tail latency later in Section~\ref{sec:eval}. 

\begin{figure}[t]
	\centering
	\includegraphics[width=\columnwidth]{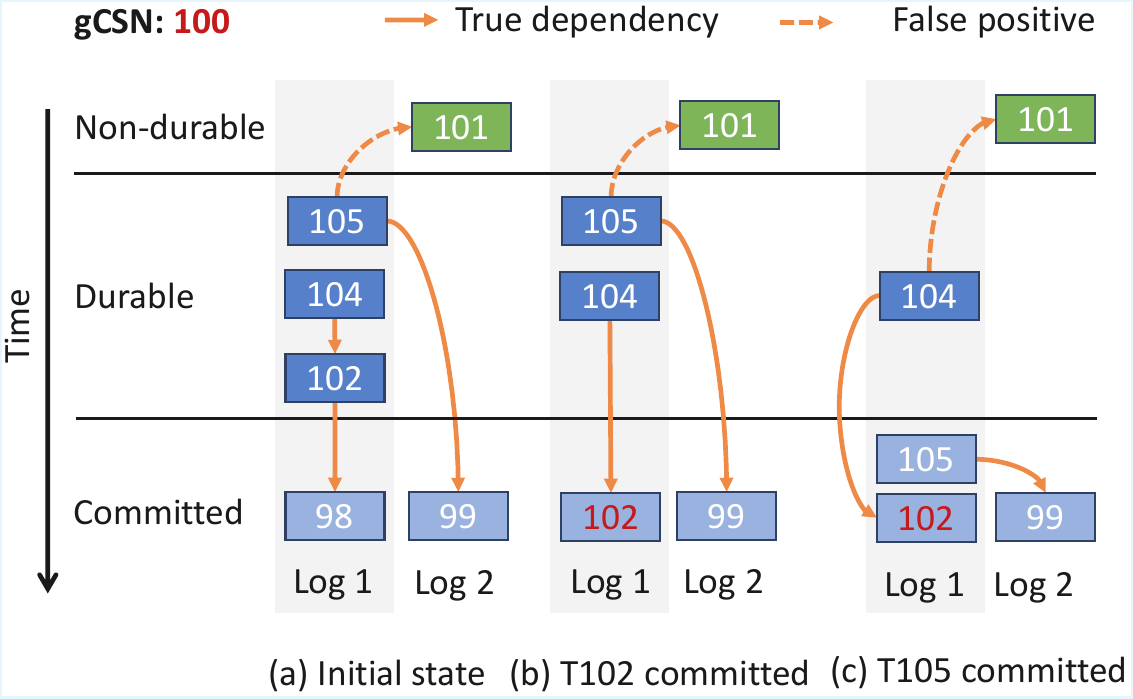}
	\caption{Pathological cases for record-level dependency tracking. 
  (a) Initially, T102, T104 and T105 are durable and the global commit sequence number (gCSN) is 100.
	(b) T102 commits as its maximum dependent sequence number [DSN, 98] is smaller than gCSN. 
	(c) T104 can not commit with a DSN [102] greater than gCSN, 
  but T105 is allowed to commit since its latest dependent transaction [99] has already committed.}
	\label{fig:worst-case}
\end{figure}

\textbf{Analysis and Limitations.}
Record-level tracking reduces more false positives than prior approaches, but do not completely eliminate them. 
Figure~\ref{fig:worst-case} reasons about the merit and limitations of our approach when the system starts with an initial state where 
$T99$ and $T98$ have committed and the commit sequence number is 100.
In Figure~\ref{fig:worst-case}(a), since $T102$ only depends on $T98$, its DSN is 98, which is less than the commit sequence number (100). 
This means that all the dependencies of $T10$2 have committed and since $T102$'s own log records are durable, $T102$ can safely commit. 
In Figure~\ref{fig:worst-case}(b) $T104$ only depends on $T102$, which has committed. 
However, the committed sequence number is less than 102, which means that not all the transactions with a commit sequence number lower than 102 are persistent (e.g., $T101$). 
As a result, $T104$ can not commit and it remains to be explored how to eliminates such false positive dependencies completely with low overhead. 
As we show in later in Section~\ref{sec:eval}, however, the impact of such pathological cases is negligible in a wide range of representative workloads. 

\subsection{Checkpointing and Recovery} 
\label{subsec:delta}
As log objects accumulates, it is beneficial to reduce recovery time by checkpointing in-memory data to object storage. 
\name inherits ERMIA's checkpointing approach~\cite{ERMIA} as the aforementioned restricted decentralized logging and record-level dependency tracking techniques do not affect checkpointing is performed. 
We describe how it works for completeness here. 
The main goal for an memory-optimized system like \name and ERMIA is to store a consistent snapshot of the in-memory data.  
The checkpointing thread first obtains a transaction begin timestamp $t$ by reading the central timestamp counter, and then scans the indirection array of each table to store the latest version that is committed before $t$. 
Thanks to MVCC, checkpointing does not block forward processing and a checkpoint will represent the state of the database as of the beginning of the checkpoint. 
The log can then be truncated by deleting S3 objects that cover the checkpointed data. 

The recovery protocol also remains similar to common redo-only logging designs~\cite{Silo,NVM-Dlog,QueryFresh,Cicada,ERMIA} where multiple threads can proceed in parallel to scan the logs and race to install the latest versions. 
The only minor adaptation needed is to ensure transactions with a DSN greater than the global durable timestamp are preserved during recovery. 
For example, it was sufficient to replay the log until $T100$ in Figure~\ref{fig:worst-case} under transaction-level dependency tracking. 
However, under record-level dependency tracking \name should ensure that transactions $T102$ and $T105$ survive after a crash. 
To facilitate this, we embed in each log record the transaction's DSN, and replay all log records whose dependent transactions have committed (i.e., with a commit timestamp greater than the log record's DSN). 

Finally, most prior redo-only logging designs for main-memory systems store in each log record the entire new version, including fields that are not updated.  
This adds log data volume, leading to more S3 append requests. 
\name stores only the delta (i.e., updated fields) in the log to reduce log data volume, but keep full versions in memory to avoid record reconstruction cost during forward processing. 
While this optimization is highly workload-dependent, as we show later in Section~\ref{sec:eval}, storing only the delta can save as much as two-thirds of data volume in TPC-C. 

\subsection{Discussions}
\label{subsec:discuss}
Restricted decentralized logging strikes a balance between contention avoidance (per-thread logging) and log buffer flush frequency. 
An alternative design is to still dedicate a private log buffer per thread, but allocate these log buffers in contiguous memory such that they can be flushed in one I/O. 
This would keep the full benefit of decentralized logging but require coordination among threads, e.g., to cope with cases where threads fill out their log buffers in different speeds which can delay the flush. 
We therefore opt for limited sharing within a group of threads. 
As mentioned in Section~\ref{subsec:rdl}, log buffer size plays an important role and is set based on current latency characteristics of \express; 
tuning it at runtime automatically is a promising future direction. 

\name can provide stronger availability than most alternatives. 
Compute-local SSDs are ephemeral, 
and so one must store data permanently in a disaggregated storage service that will survive compute restarts, such as EBS. 
However, its availability is limited to a single AZ. 
Although data in EBS can be periodically snapshot and stored in other AZs, the storage-level snapshot boundary may not coincide with that of the DBMS's, leaving the DBMS largely to handle replication. 
Moreover, 
Compared to \std, \express provides the desirable low latency by trading off high availability guarantees. 
Therefore, systems like \name may need to replicate data manually to reach the same level of high availability \std. 
\name allows the application to provide a list of S3 buckets for replication. 
Upon group commit, we issue multiple asynchronous S3 append requests to all the buckets and start commit after all or a majority of requests have finished. 

Finally, our focus in this paper has been adapting memory-optimized OLTP engines without buffer pools to use \express. 
However, both techniques are applicable beyond such engines and can be leveraged by buffer pool-based systems~\cite{LeanStore,Umbra} since their logging subsystems largely follow the same designs. 
For example, one may apply restricted decentralized logging on top of the distributed logging design in LeanStore~\cite{LeanStore}. 
We leave such explorations to future work.

%% file: 5-evaluation.tex
\section{Evaluation}
\label{sec:eval}
We have shown the effect of restricted decentralized logging and record-level dependency tracking over a set of microbenchmarks earlier. 
Now we expand our evaluation of \name with a wider range of baselines and benchmarks to demonstrate the following:
\begin{itemize}[leftmargin=*]
\item \name can easily outperform S3 standard in terms of latency and throughput across a variety of benchmarks.
\item \name can achieve comparable average and tail commit latency, if not lower, compared to baselines using block storage such as EBS while sustaining high throughput. 
\item Techniques in \name are also proved useful for other storage backends, including on prem and cloud block storage.  
\end{itemize}

\subsection{Experimental Setup}
\textbf{Hardware.}
We perform experiments on a dedicated Amazon EC2 \textsf{c6in.8xlarge} instance over a variety of storage backends. 
The instance is equipped with an Intel Xeon Platinum 8375C CPU clocked at 2.9GHz (up to 3.5GHz with turboboost) and 64GB of memory. 
Although the CPU itself has 32 cores (64 hyperthreads), the instance exposes 32 vCPUs (hyperthreads) over 16 physical cores. 

In addition to \std and \express, we use EBS gp3 and io2 as baselines. 
Both gp3 and io2 are provisioned with 8,000 IOPS. 
The gp3 volume offers up to 2,000Mbps bandwidth while the io2 volume offers unlimited bandwidth.
The instance supports 50Gbps of network bandwidth and 25Gbps of EBS bandwidth, and has no local NVMe SSD to lower overall cost of ownership, so all persistent data is either written to EBS or S3. 

\textbf{Software.}
The server runs Ubuntu 24.04.2 LTS and all code is compiled using GCC 13 with all optimizations enabled. 
We base our implementation on AWS SDK for C++ (version 1.11.676)~\cite{S3SDK}. 
The original AWS SDK provides asynchronous interfaces for S3 append, but does so by using a thread pool that is opaque to the application (i.e., \name), which adds CPU scheduling overhead. 
We therefore implemented our own asynchronous S3 operations that internally spawns self-managed flusher threads to invoke the synchronous AWS S3 append interfaces. 

Each worker thread group in \name is allocated a flusher thread. 
We observe that they only take 3--5\% of CPU cycles, so they could be pinned to the same hyperthread in budget constrained scenarios. 
To ease the interpretation of our results, we run experiments under 16 worker threads, each pinned to a different core. 
Each flusher thread is pinned to a different hyperthread of the same cores used by the corresponding worker thread group. 
For baselines that use block storage (gp3 or io2) and per-thread logs, we use io\_uring to issue asynchronous I/O operations and each log (worker) has its own flusher thread which is necessary to keep up with the fill rate and sustain high throughput. 

\textbf{Experimental Variants.}
We implemented all variants on top of an improved version of ERMIA~\cite{ERMIA} by 
replacing its original centralized logging with per-thread decentralized logging with pipelined commit (Section~\ref{subsec:log}), transaction-level dependency tracking and delta-based logging (Section~\ref{subsec:delta}), and 
(2) using double buffering for log flush to overlap with forward processing, as commonly done in prior work. 
We then change the storage backend and commit protocol to build the following variants for our experiments: 
\begin{itemize}[leftmargin=*]
  \item \sthree: Baseline that directly persists log records to \std, which enables record-level dependency tracking. It still keeps the per-thread dedicated logging for best average latency, because restricted decentralized logging does not improve tail latency atop storage backends as slow as \std.  
	\item \xp: Baseline that directly persists log records to \express, without optimizations proposed for \name (i.e., naive use of \express). 
	\item \ebsg: Compared to \xp, it uses gp3 as the storage backend instead of \express. 
  \item \ebsi: Compared to \xp, it uses io2 as the storage backend. 
  \item \milliscale: Same as \xp but applies both restricted decentralized logging and record-level dependency tracking. 
\end{itemize}

Same as prior work on OLTP engines~\cite{Silo,ERMIA,AutonomousCommit,LeanStore,Umbra}, we implement \name and baselines as shared libraries. 
The application (such as various benchmarks) directly uses the engine's C++ APIs to access database records, without SQL and optimizer layers. 
All the variants inherits the same concurrency control protocols in ERMIA and support read committed, snapshot isolation and serializability. 
The choice of isolation level is orthogonal to our evaluation, and we use snapshot isolation for all experiments.  

Based on observations from Section~\ref{subsec:s3xp}, we use 512KB log buffer for variants that use per-thread logging. 
For \milliscale, we set log group size to two. 
Following analysis in Section~\ref{subsec:rdl}, since each group consists of two threads, we set log buffer is 1MB (shared by two threads) so that on average each thread has the same space as in per-thread logging.

\textbf{Benchmarks.}
We use both microbenchmarks based on YCSB~\cite{YCSB} and end-to-end TPC-C benchmarks~\cite{TPCC}. 
For YCSB, we use a database table of 30 million records. 
Each record is 80 bytes, including ten 8-byte fields, with one of them being the primary key. 
Each transaction follows a uniform or skewed (zipfian with theta=0.99) distribution to point read or update (read-modify-write) ten records through a Masstree~\cite{Masstree} index built over the table's primary key column; 
the update operation reads the entire chosen record and modifies a random field in it. 
We test two transaction mix profiles, YCSB-A and YCSB-B, which respectively issues 50\%/50\% read/update and 95\%/5\% read/update transactions.
For TPC-C, we use a database of 100 warehouses, and let each transaction randomly chooses a home warehouse to access. 
To focus on logging, we keep all table data in memory, so that all the traffic going to the storage backend is incurred by flushing log buffers.

We report the throughput, average commit latency and tail commit latency for all workloads. 
Note that for latency, we only measure the commit delay after forward processing. 
This allows us to fairly compare commit latency across all transaction types. 
Each experiment is run for 10 seconds and repeated for three times; 
the variance across runs is minuscule and so we report the average. 

\subsection{Microbenchmark Scalability} 
Our first set of experiments evaluate the throughput and commit latency under the YCSB-A and YCSB-B microbenchmarks. 

\begin{figure}[t]
	\centering
	\includegraphics[width=\columnwidth]{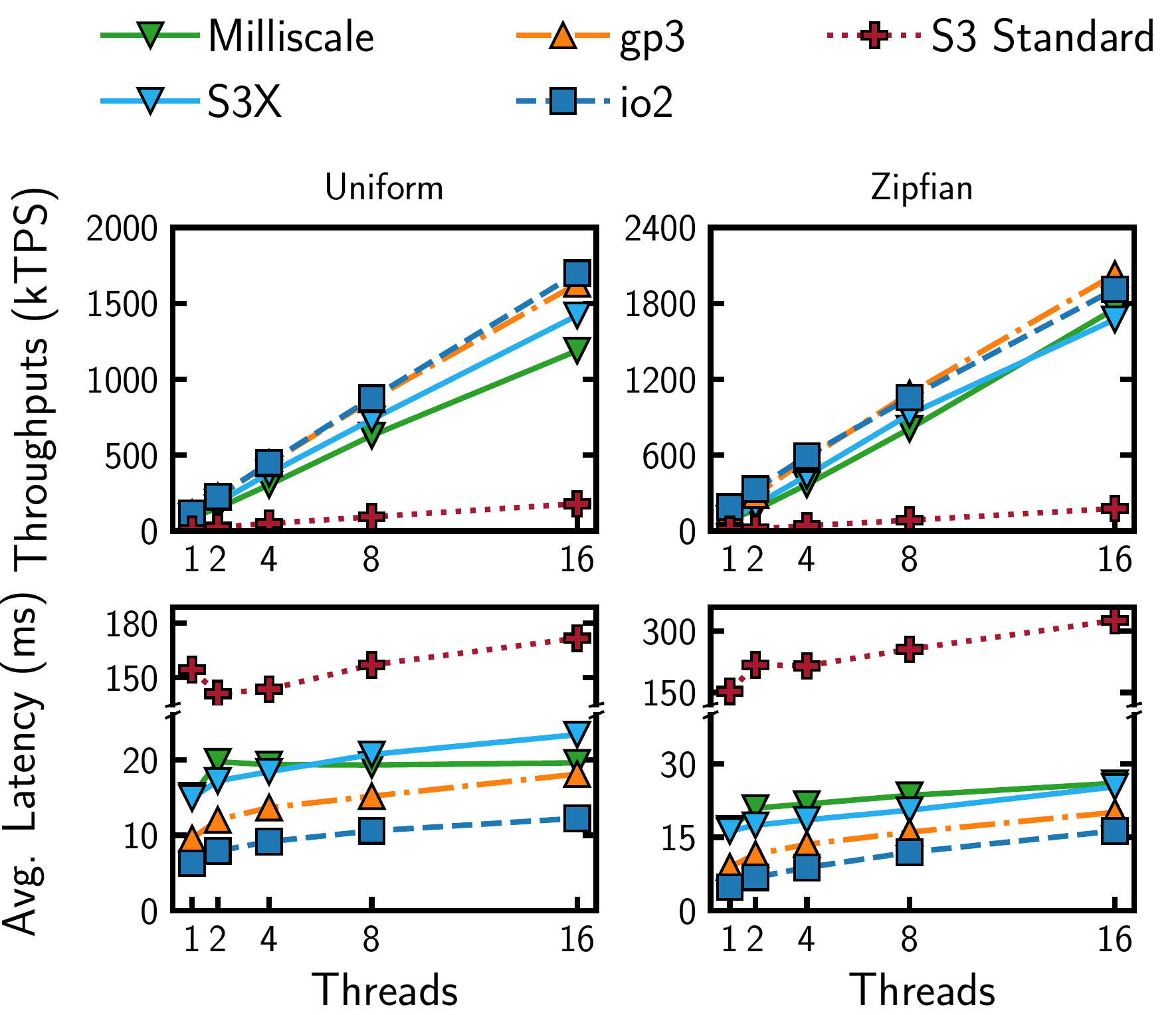}
	\caption{Throughput and average latency of YCSB-A (50\% read, 50\% write) under uniform (left) and zipfian (right) distributions.
  All but \sthree scale. 
  \milliscale sustains high throughput as baselines and achieves latency similar to that of using block storage backends. }
\label{fig:ycsba}
\end{figure}

\begin{figure}[t]
	\centering
	\includegraphics[width=\columnwidth]{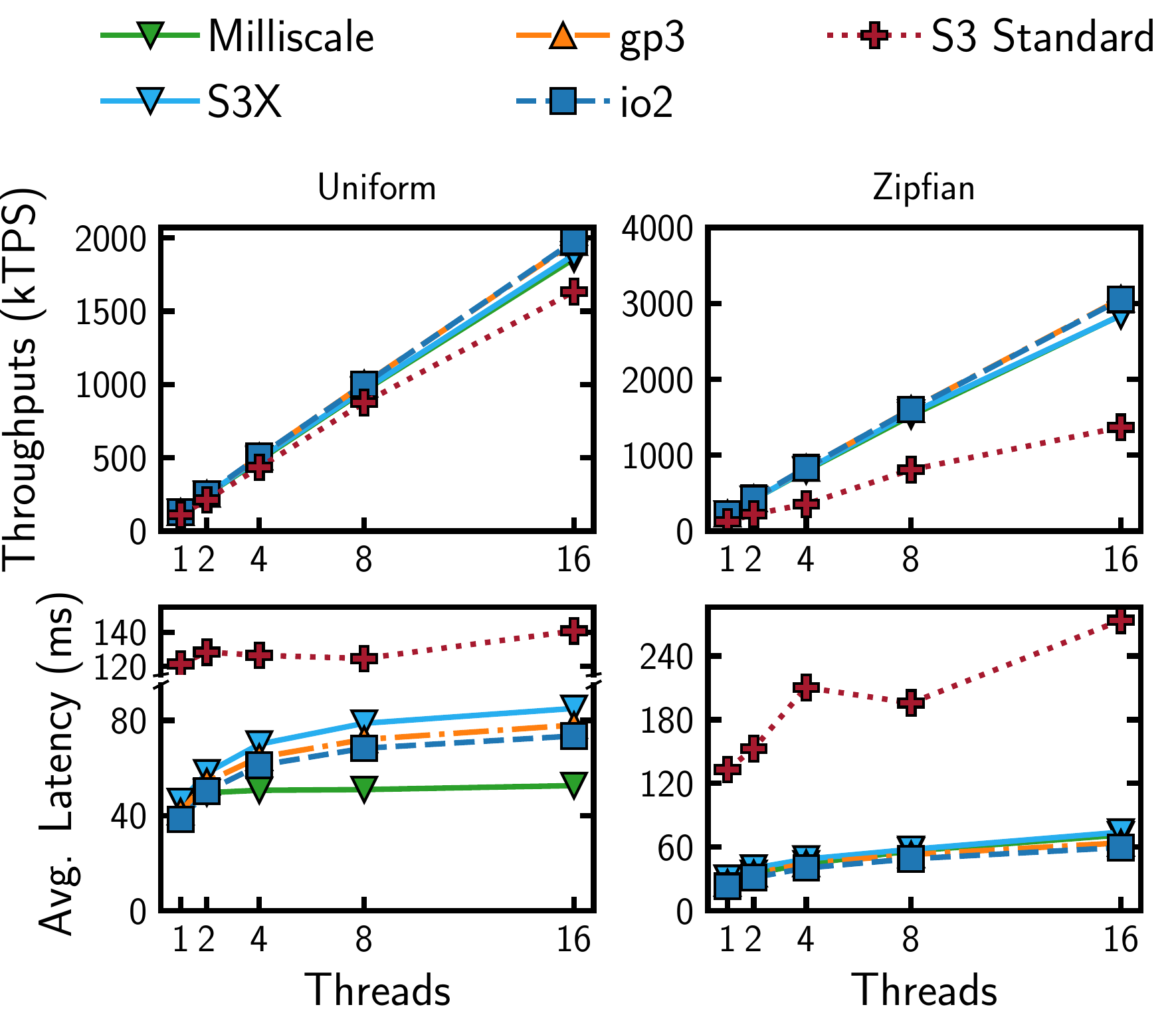}
	\caption{Throughput and average latency of YCSB-B (95\% read, 5\% update) under uniform (left) and zipfian (right) distributions.
  With more (in-memory) read operations \milliscale achieves lower latency than block storage, thanks to record-level dependency tracking.
  }
\label{fig:ycsbb}
\end{figure}

\textbf{Throughput.} 
As shown in Figure~\ref{fig:ycsba}(top),  all variants except \sthree scale well as thread count increases under uniform distribution, thanks to pipelined commit and double buffering that are able to hide storage latency. 
The same trend is seen under the skewed distributed in Figure~\ref{fig:ycsba}(top).  
\sthree's high latency well exceeds the workload's log buffer fill rate, forbidding it to scale. 
Compared to YCSB-A, YCSB-B includes more reads which leads to lower log buffer fill rate as a result of fewer updates (hence lighter log traffic). 
As a result, double buffering can hide most log flush delays and leading to higher throughput shown in Figure~\ref{fig:ycsbb}(top).  

\textbf{Average Latency.} 
Next we explore how average commit latency changes under varying thread count; 
we explore tail latency later. 
Under the uniform distribution, \milliscale maintains nearly constant average latency (similar to single-threaded cases) across all thread counts as shown in Figure~\ref{fig:ycsba}(bottom). 
Under 16 threads, \milliscale exhibits 28.4\% lower average latency in YCSB-A than \xp; 
the number for YCSB-B in Figure~\ref{fig:ycsbb}(bottom) is 42\% lower. 
These improvements are mainly a result of \name's record-level dependency tracking which is more fine grained than prior approaches. 
Note that restricted decentralized logging can slightly increase latency due to more than one thread sharing the same log buffer. 
Compared to the default sharing ratio of 2:1, using a 1:1 ratio allows YCSB-A to achieves a 35\% latency reduction under 16 threads. 
However, as we see later, restricted decentralized logging can more significantly reduce tail latency. 
YCSB-B benefits more from record-level dependency tracking: 
the workload itself features fewer writes and write transactions conflict less often under a uniform distribution, leading to fewer true dependencies that \name can effectively handle. 

Skewed workloads under the zipfian distribution present more dependencies across transactions, especially true dependencies. 
Yet record-level tracking is most effective for avoiding blocking transactions to commit due to false dependencies. 
Therefore, under YCSB-A in Figure~\ref{fig:ycsba}(right) the average latency of \milliscale is close to that of \xp across all thread counts. 
YCSB-B in Figure~\ref{fig:ycsbb}(right) exhibits similar results, but with negligible gaps between \xp and EBS variants because the amount of writes is only 5\%. 

\subsection{Tail Latency}
\begin{figure}[t]
	\centering
	\includegraphics[width=0.75\columnwidth]{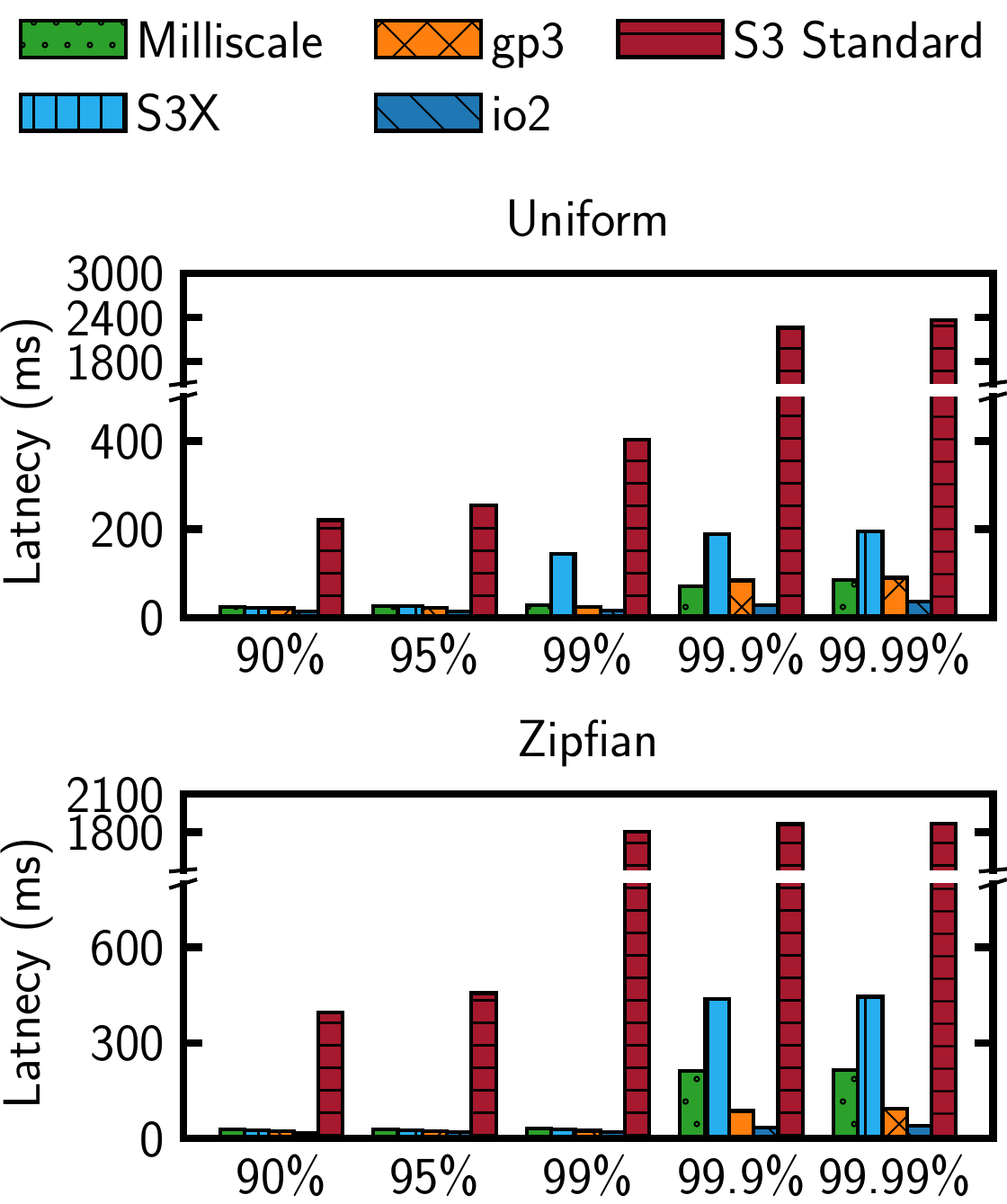}
	\caption{YCSB-A tail latency under 16 threads.
  \milliscale achieves comparable latency to \ebsg's under uniform distribution. 
  Skewed workloads (bottom) present more true dependencies and increases latency for \milliscale, which however still exhibits over 50\% lower latency at 99.9\% and 99.99\%.}
	\label{fig:ycsb_a_tail_latency}
\end{figure}

\begin{figure}[t]
	\centering
	\includegraphics[width=0.8\columnwidth]{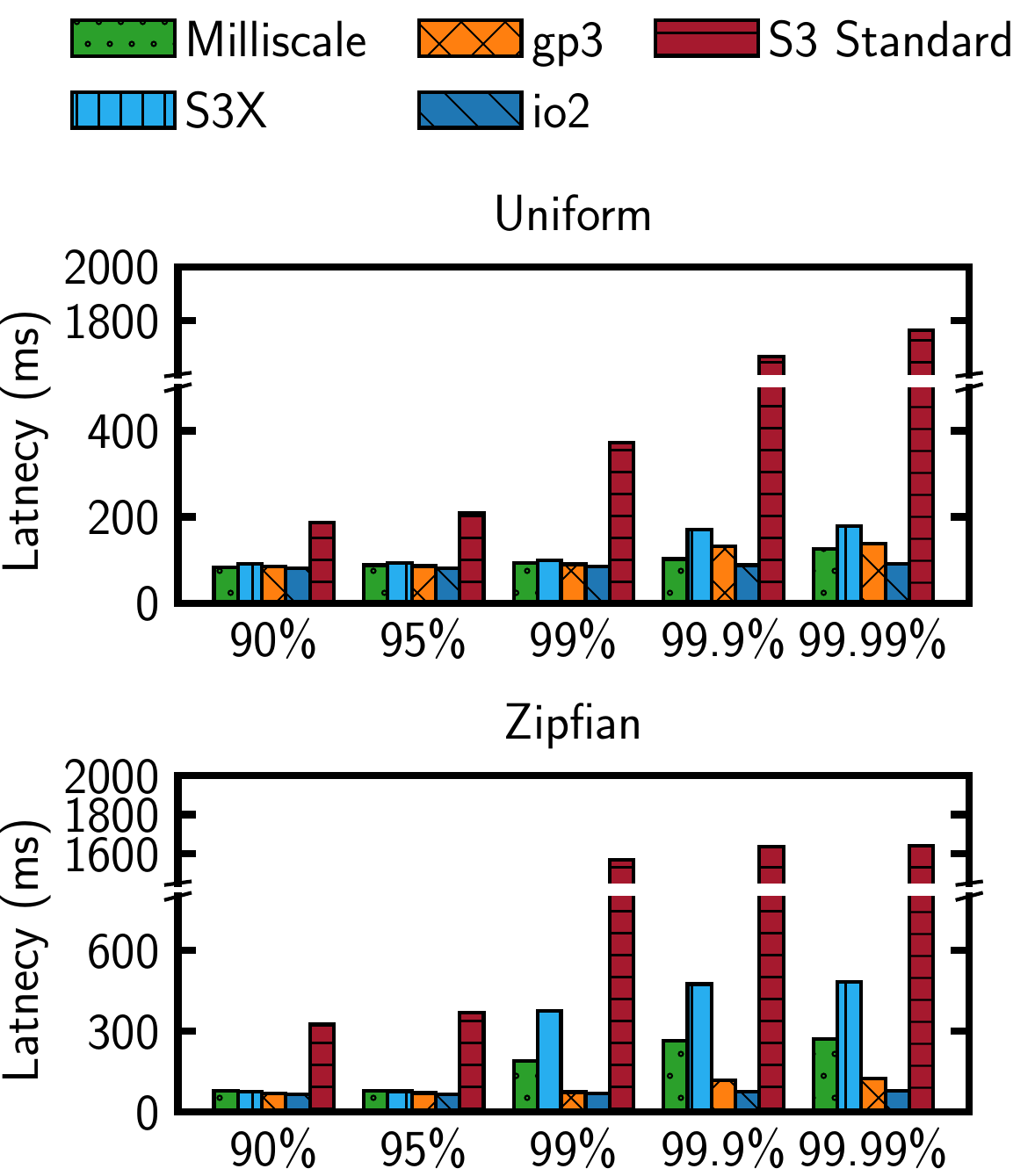}
	\caption{YCSB-B tail latency under 16 threads and uniform (top) and skewed (bottom) accesses.}
	\label{fig:ycsb_b_tail_latency}
\end{figure}

In this experiment, we fix the number of threads to 16 and perform YCSB-A and YCSB-B workloads under both uniform and zipfian distributions.  
Figure~\ref{fig:ycsb_a_tail_latency} shows the commit latency for YCSB-A under different percentiles. 
Here, \sthree and \ebsi respectively exhibits the highest and lowest tail latency across all percentiles.  
They represent two extremes: cheap but slow vs. fast but expensive. 
Other variants attempt to strike a balance between these extremes. 
\milliscale and \ebsg achieve similar tail latency profiles under uniform distribution, but the former starts to exhibit higher latency than \ebsg under skewed accesses. 
We attribute the reason to more true dependencies under skewed workloads.
Nevertheless, \milliscale still significantly reduces 99.9 percentile latency than \xp by 62.4\%/51.8\% under uniform/zipfian distributions. 
The numbers for 99.99 percentile tail latency are 56.2\%/51.6\%.  
Results of running YCSB-B as shown in Figure~\ref{fig:ycsb_b_tail_latency} give similar trends, so we do not repeat here.


\subsection{Ablation Study}
\begin{figure}[t]
  \centering
  \includegraphics[width=\columnwidth]{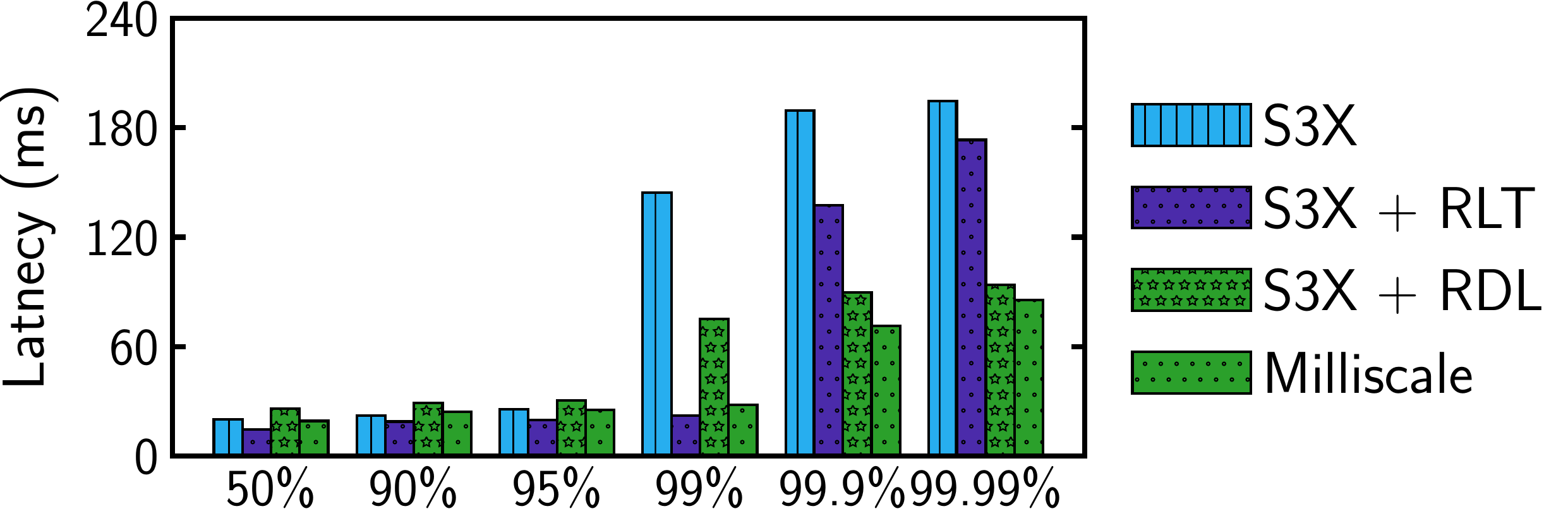}
  \caption{Effect of record-level dependency tracking (\xp + RLT), restricted decentralized logging (\xp + RDL) and their effects combined (\milliscale) 
  under 16 threads and YCSB-A.}
  \label{fig:ablation}
\end{figure}

Section~\ref{sec:design} has analyzed the effectiveness of restricted decentralized logging and record-level dependency tracking for average commit latency. 
Now we expand the evaluation to tail latency using YCSB-A.  
We start with \xp, and then separately evaluate each technique under different percentiles.  
Figure~\ref{fig:ablation} shows the result. 
All variants exhibit reasonably low latency until 95 percentile. 
As expected, restricted decentralized logging (\textsf{S3X + RDL} in the figure) exhibits slightly higher tail latency at 50 and 99 percentiles than record-level dependency tracking (\textsf{S3X + RLT}) due to slightly increased log buffer contention from threads in the same group that share a log buffer. 
Starting from 99 percentile, \xp's latency grows significantly by over 5$\times$.
Record-level dependency tracking helps stretch the trend to 99 percentile and remains helpful at 99.9 and 99.99 percentiles. 
At 99.9\% and 99.99\% restricted decentralized logging is more effective than record-level dependency tracking. 
We observe the reason is that with fewer log buffers, the workload inherently presents fewer cross-log dependencies that must be resolved before committing (e.g., $T105$ as identified in Figure~\ref{fig:worst-case}).

\begin{figure}[t]
	\centering
	\includegraphics[width=\columnwidth]{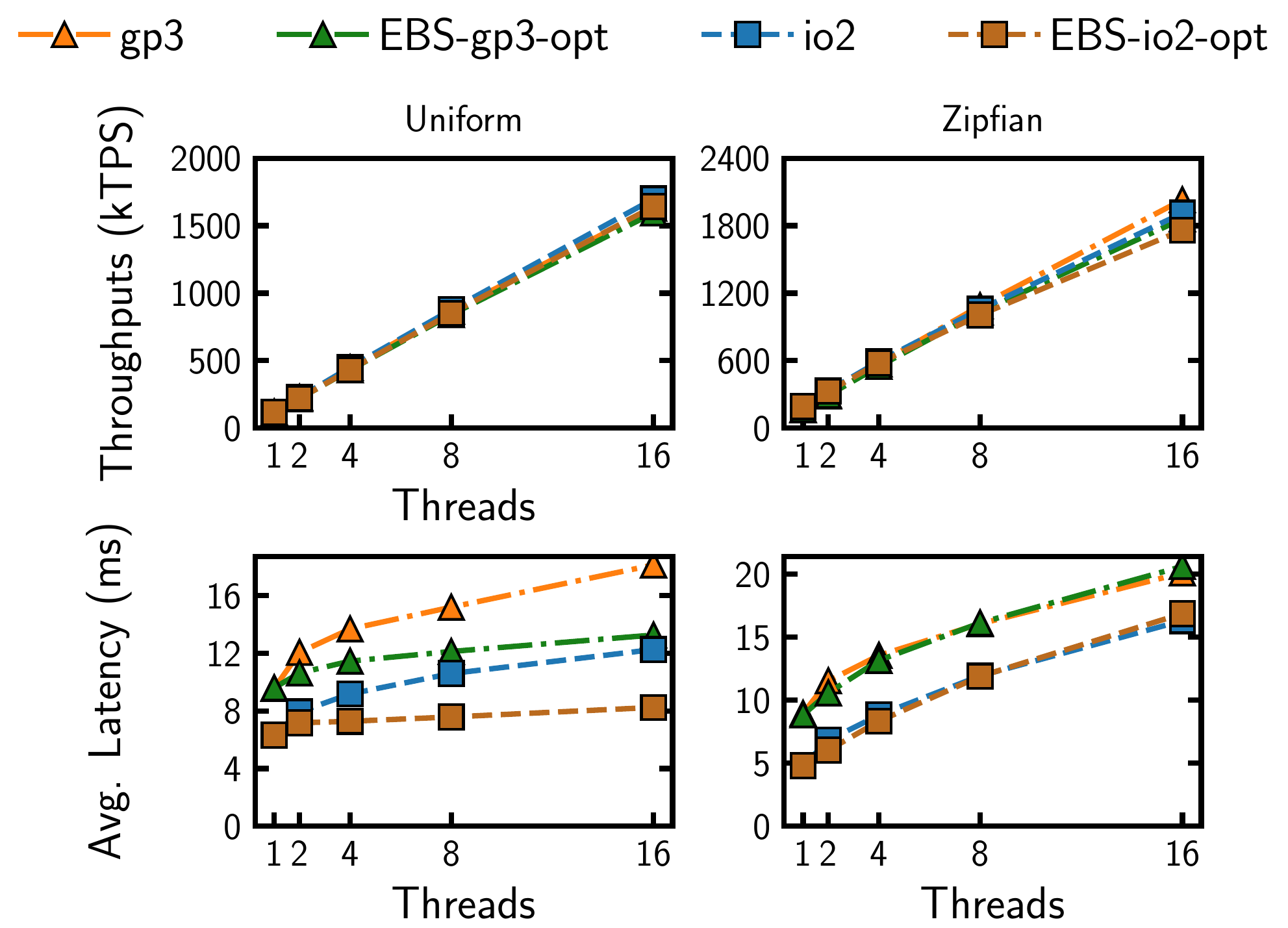}
	\caption{Throughput and average latency of YCSB-A when using EBS gp3/io2.
  \name optimizations (\textsf{*-opt}) is mostly useful for under the uniform distribution for block storage.}
	\label{fig:ebs_ycsba}
\end{figure}

\subsection{Effect on Block Storage}
\begin{figure}[t]
	\centering
	\includegraphics[width=0.85\columnwidth]{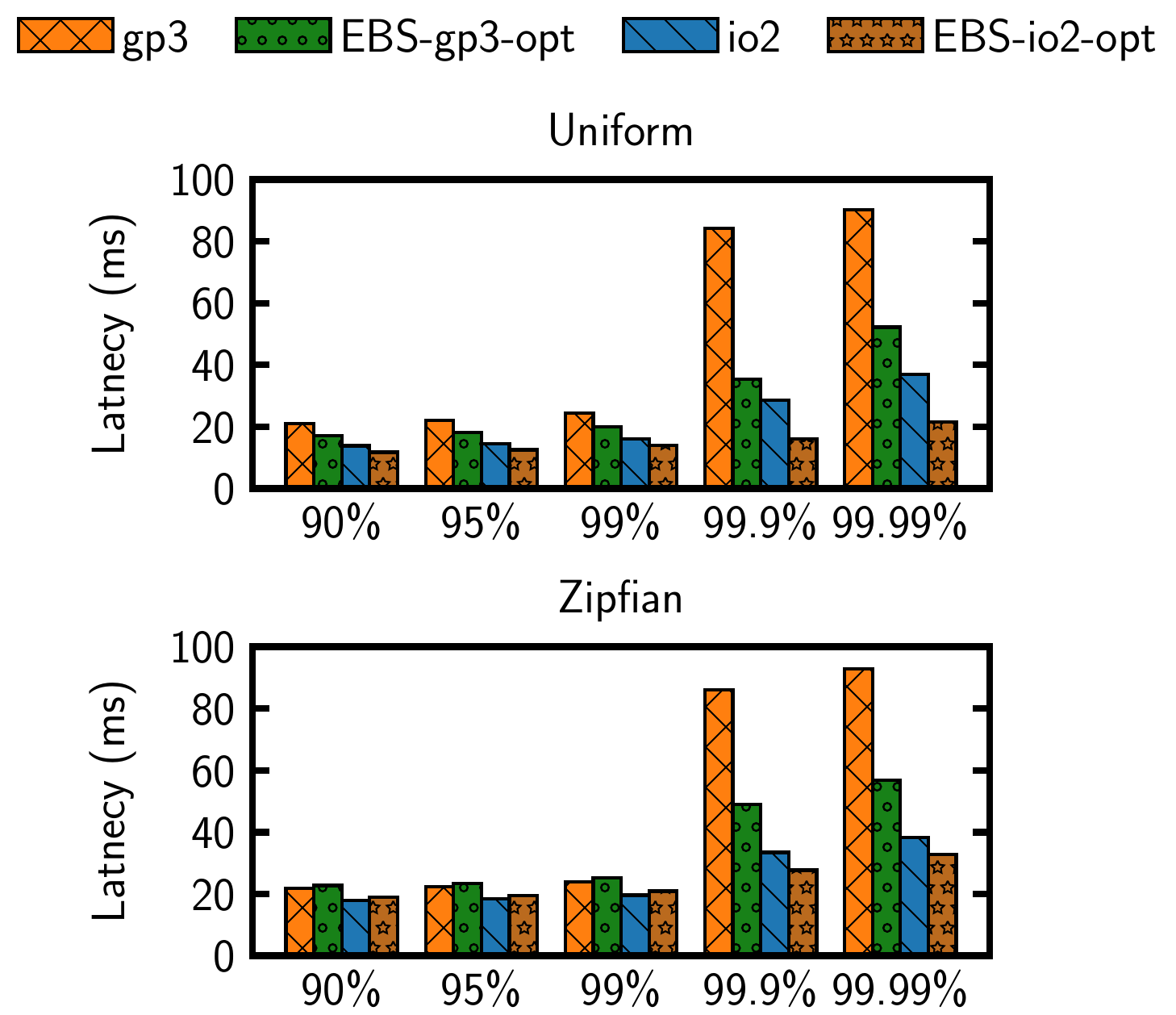}
	\caption{Tail latency of YCSB-A (16 threads) on block storage backends with vs. without \name optimizations.} 
	\label{fig:ebs_tail}
\end{figure}

As mentioned earlier, the usefulness techniques in \name are not limited to object storage and can be applied to EBS or even on prem SSDs.  
We apply restricted decentralized logging (with the same 2:1 sharing ratio) and record-level dependency tracking on top of \ebsg and \ebsi. 
Figures~\ref{fig:ebs_ycsba} and~\ref{fig:ebs_tail} respectively show throughput/average latency and tail latency behaviors under YCSB-A. 
With optimizations from \name, the average latency for \ebsg and \ebsi under 16 threads decreased by 26.8\% and 32.8\%, respectively. 
The numbers for 99.99 percentile latency are 42.2\% and 41.6\%, separately.
Similar to previous results, under the skewed zipfian distribution (where transactions exhibit more dependencies), average latency and throughput show no significant drop, and tail latency under 16 threads decreased by 39.0\% for \ebsg and 14.4\% for \ebsi.

Overall, these results indicate that \name techniques are useful for faster storage backends, but are more needed and effective for higher latency counterparts, such as \express.

\subsection{End-to-End TPC-C Results}
\begin{figure}[t]
  \centering
  \includegraphics[width=0.8\columnwidth]{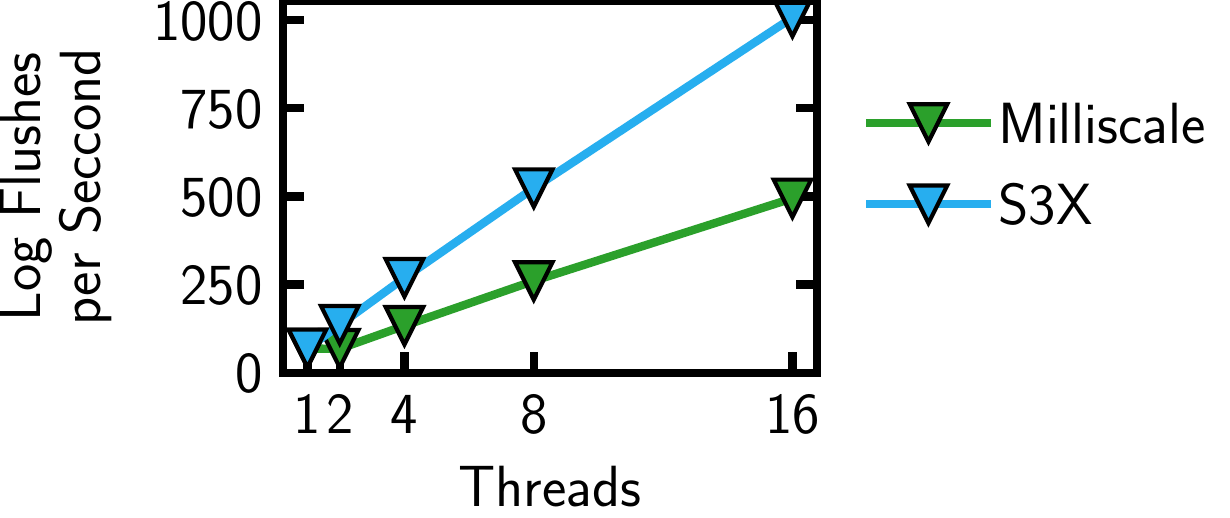}
  \caption{Number of S3 append requests per second under TPC-C using restricted decentralized logging with a 2:1 sharing ratio (1MB log buffer shared by two threads) vs. per-thread logging (512KB buffer per thread).} 
  \label{fig:tpcc-requests}
\end{figure}

\begin{figure}[t]
	\centering
	\includegraphics[width=\columnwidth]{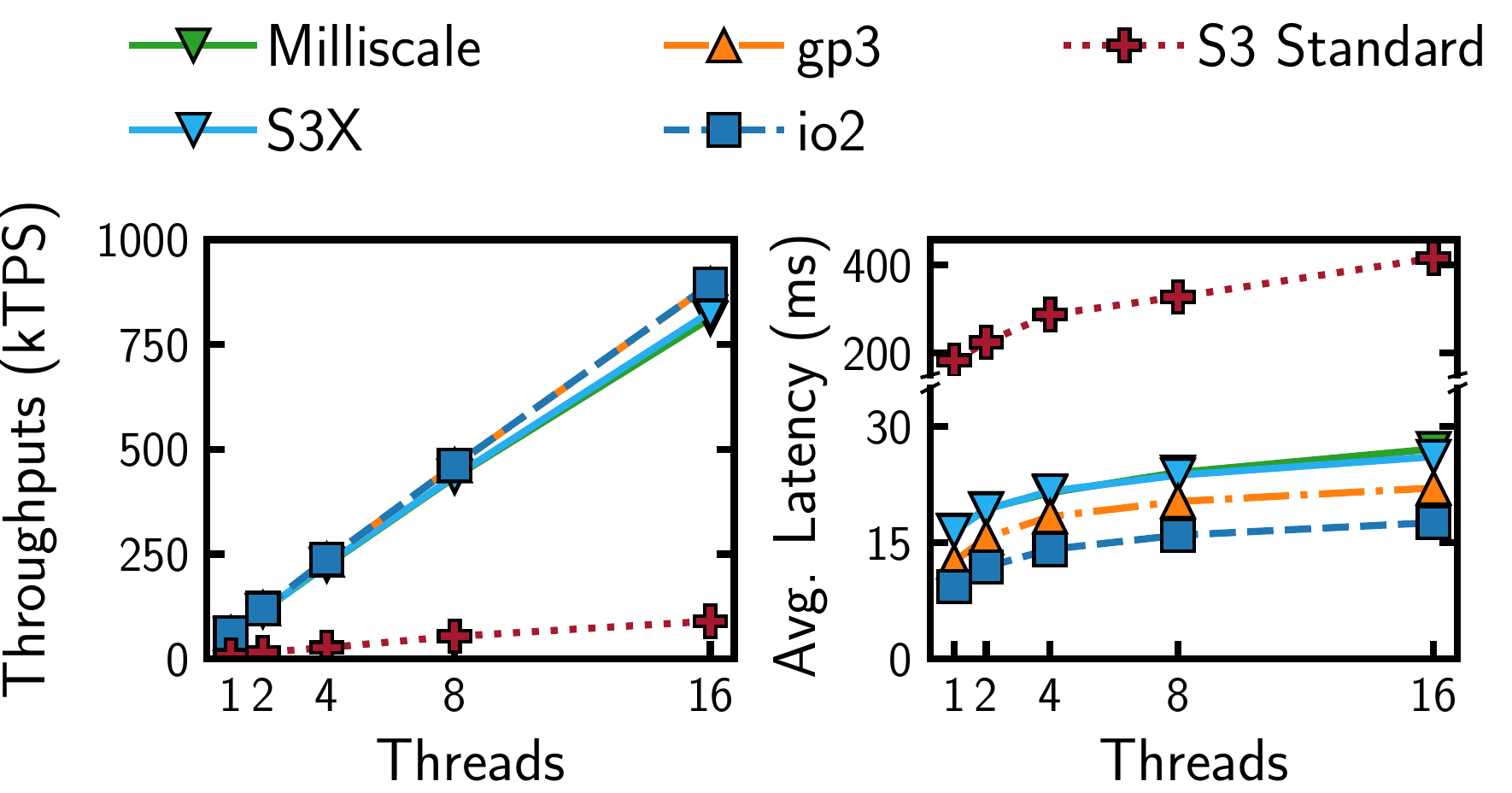}
	\caption{TPC-C throughput and average latency.}
	\label{fig:tpcc-scal}
\end{figure}

\begin{figure}[t]
	\centering
	\includegraphics[width=\columnwidth]{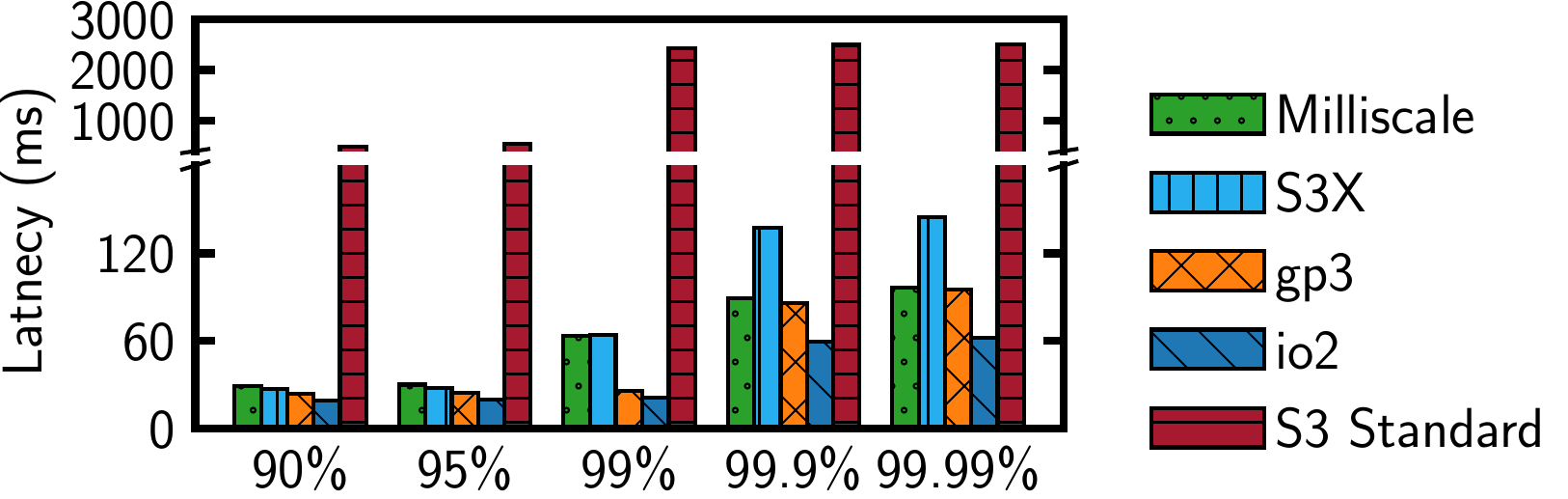}
	\caption{TPC-C tail latency under 16 threads.}
	\label{fig:tpcc-tail}
\end{figure}

Our final experiment tests the performance of \name using end-to-end TPC-C. 
As mentioned earlier, we use a database of 100 warehouses and let each worker thread randomly choose its home warehouse. 
This allows to stress the system with more cross-log dependencies and increases contention between transactions.
With a 2:1 sharing ratio, restricted decentralized logging in \milliscale cuts the number of S3 append requests by $\sim$50\% across different scales in Figure~\ref{fig:tpcc-requests}.  
As show in Figure~\ref{fig:tpcc-scal}, \milliscale outperforms the baseline \sthree significantly in terms of both throughput and latency. 
\milliscale's average latency is 3.8\% higher than \xp and 23\% higher than \ebsg under 16 threads. 
Under 99 percentile, both \xp and \milliscale exhibit much higher latency than \ebsg and \ebsi, which we attribute to the difference of each storage backend's performance characteristics: 
\express's tail latency starts to spike at 99 percentile, while for \ebsg and \ebsi this starts at 99.9 percentile. 
Nevertheless, as shown in Figure~\ref{fig:tpcc-tail}, similar to earlier microbenchmark results, \milliscale effectively lowers latency compared to \xp and achieves similar to \ebsg's latency at 99.9 and 99.99 percentiles. 

%% file: 6-related-work.tex
\section{Related Work}
\label{sec:related}

Our work is most related to prior work that leverages object storage for DBMSs and optimizes write-ahead logging. 

\textbf{DBMSs over Object Storage.}
The earliest attempt (to the best of our knowledge) of using S3 for DBMS was done by Brantner et al.~\cite{S3DB}. 
The system was based on \std and had to deal with many drawbacks of S3, in particular weak consistency, lack of append support and very high latency (100ms-level).
These led to solutions that are not fully ACID compliant and logging was offloaded to a separate message queue implementation. 
The high end-to-end transaction latency (seconds-level) has discouraged these approaches. 
As Section~\ref{sec:bg} describes, these issues are not largely resolved (strong consistency and append support) or mitigated (latency) by \express, which allowed us to build \name.  
Delta Lake~\cite{DeltaLake} manages transactions over \std for data lake applications. 
To bridge the speed gap between memory and \std, it relies on heavy caching using compute-side SSDs. 
Other approaches in this space mostly focused on OLAP or hybrid workloads over \std. 
Colibri~\cite{Colibri} proposes a hybrid architecture for OLTP and OLAP workloads to leverage cloud storage and separate hot and cold pages.  
AnyBlob~\cite{AnyBlob} presents an io\_uring-based download manager that asynchronously handles concurrent \textsf{S3 GET} requests, enabling DBMSs to saturate fast data center networks for read-heavy OLAP workloads while minimizing CPU overhead. 
LiquidCache~\cite{LiquidCache} deploys a cache server between compute nodes and object storage. 
While data is stored in object storage as Parquet files, the cache server uses a custom in-memory format optimized for fast filter evaluation. 
PushdownDB~\cite{PushdownDB} uses the (now end-of-life) S3 Select~\cite{S3Select} feature to push filter, projection, and aggregation operations directly to S3 storage nodes, reducing network traffic between storage and compute for OLAP workloads.
FlexPushdownDB~\cite{FlexPushdownDB} extends PushdownDB with a hybrid architecture that combines caching with computation pushdown at fine granularity.
Compared to prior work, \name targets a different storage backend (\express) and focuses on optimizing write-ahead logging for lower commit latency.  

\textbf{Logging Optimizations.}
\name is built on top of several important prior ideas for scalable logging.  
Aether~\cite{Aether} evaluated and showed the potential of early lock release and log flush pipelining, which allowed subsequent systems to improve transaction throughput to the same level of asynchronous commit, yet without sacrificing correctness. 
However, the inherent centralized logging bottleneck still demands decentralized logging~\cite{AetherVLDBJ,SiloR,TaurusLog}, which later became a standard design decision in memory-optimized OLTP engines~\cite{Cicada,Silo,LeanStore,Umbra}. 
A major drawback in traditional decentralized logging is that it trades transaction commit latency for throughput. 
Cross-log dependencies are a major source, and various prior approaches have been proposed to mitigate their impact.  
For example, LeanStore uses remote flush avoidance~\cite{RethinkLog} to skip unnecessary log buffer flushes if a transaction does not depend on other logs. 
Autonomous commit~\cite{AutonomousCommit} parallelizes small log flushes across worker threads and proactively inserts dummy transactions to advance GSNs to mitigate the impact of stragglers.
Border-Collie~\cite{BorderCollie} proposes wait-free algorithms to efficiently find the upper bound logical clock position for transactions to safely commit. 
ELEDA~\cite{ELEDA} fill gaps in LSNs across logs so that the globally durable LSN marker can be advanced timely. 
Compared to these approaches, \name tracks and resolves dependencies at the finer record granularity while maintaining low overhead.

%% file: 7-summary.tex
\section{Summary}
\label{sec:summary}
We have explored the feasibility of leveraging recent low-latency, mutable object storage services as represented by Amazon \express for OLTP write-ahead logging. 
While traditional decentralized logging approaches can deliver high throughput, they present high transaction commit latency (especially tail latency) over \express. 
Based on \express's performance characteristics, we propose \name, a memory-optimized OLTP engine that uses \express as write-ahead logging storage to optimize commit latency.  
\name consists of two key techniques---restricted decentralized logging and record-level dependency tracking---that respectively reduces log flush frequency without imposing higher log data transfer time and avoids false positive dependencies commonly found in prior work.  
These techniques can also be applied in other systems or storage backends. 
Evaluation using both microbenchmarks and end-to-end TPC-C showed that \name can effectively reduce both average and tail latency, while sustaining high throughput.